\documentclass[a4paper,twocolumn]{emulateapj}
\usepackage{amstext}
\usepackage{apjfonts}
\usepackage{amsmath}
\usepackage{amssymb}
\usepackage{xcolor,xspace}
\usepackage{multirow,threeparttable,tabularx}
\usepackage[colorlinks,breaklinks,linkcolor=blue,citecolor=blue]{hyperref}
\usepackage{aas_macros}
\usepackage[all]{hypcap}

\newcommand{\appropto}{\mathrel{\vcenter{
  \offinterlineskip\halign{\hfil$##$\cr
    \propto\cr\noalign{\kern2pt}\sim\cr\noalign{\kern-2pt}}}}}

\begin{document}

\shorttitle{Semi-Relativistic Hypervelocity Stars}

\shortauthors{Guillochon, Loeb}

\title{The Fastest Unbound Stars in the Universe}

\author{James Guillochon\altaffilmark{1,2} and Abraham Loeb\altaffilmark{1}}
\altaffiltext{1}{Harvard-Smithsonian Center for Astrophysics, The Institute for Theory and
Computation, 60 Garden Street, Cambridge, MA 02138, USA}
\altaffiltext{2}{Einstein Fellow}

\email{jguillochon@cfa.harvard.edu}

\begin{abstract} 
The discovery of hypervelocity stars (HVS) leaving our galaxy with speeds of nearly $10^{3}$ km~s$^{-1}$ has provided strong evidence toward the existence of a massive compact object at the galaxy's center. HVS ejected via the disruption of stellar binaries can occasionally yield a star with $v_{\infty} \lesssim 10^4$ km~s$^{-1}$, here we show that this mechanism can be extended to massive black hole (MBH) mergers, where the secondary star is replaced by a MBH with mass $M_2 \gtrsim 10^5 M_{\odot}$. We find that stars that are originally bound to the secondary MBH are frequently ejected with $v_{\infty} > 10^4$ km~s$^{-1}$, and occasionally with velocities $\sim 10^5$ km~s$^{-1}$ (one third the speed of light), for this reason we refer to stars ejected from these systems as ``semi-relativistic'' hypervelocity stars (SHS). Bound to no galaxy, the velocities of these stars are so great that they can cross a significant fraction of the observable universe in the time since their ejection (several Gpc). We demonstrate that if a significant fraction of MBH mergers undergo a phase in which their orbital eccentricity is $\gtrsim 0.5$ and their periapse distance is tens of the primary's Schwarzschild radius, the space density of fast-moving ($v_{\infty} > 10^{4}$~km~s$^{-1}$) SHS may be as large as $10^{3}$ Mpc$^{-3}$. Hundreds of the SHS will be giant stars that could be detected by future all-sky infrared surveys such as {\it WFIRST} or {\it Euclid} and proper motion surveys such as {\it LSST}, with spectroscopic follow-up being possible with {\it JWST}.
\end{abstract}


\section{Introduction}
Typical stellar velocities throughout the Milky Way are a few hundred km s$^{-1}$. However, there are particular sub-populations of stars that are found to move at greater velocities; these include the hypervelocity stars \citep[HVS; The fastest having $v \lesssim 700$ km~s$^{-1}$,][]{Brown:2005a,Brown:2011a,Brown:2014a}, runaway stars \citep[e.g.][]{Heber:2008a}, and a small fraction of compact objects, with examples being the binary white dwarf LP400-22 \citep[$v > 830$ km s$^{-1}$]{Kilic:2013a} and the kicked pulsar PSR 2224+65 \citep[$\approx 10^{3}$~km~s$^{-1}$,][]{Cordes:1993a}. While these objects are moving quickly as compared to most other stars in the galaxy, they still travel at speeds significantly below that observed for stars in our own galactic center, where the velocity of the star with the closet-known approach to the central black hole exceeds $10^{4}$ km s{$^{\smash{-1}}$ \citep{Ghez:2005a}, 3\% the speed of light.

In this paper we describe a mechanism by which binary massive black hole (BMBH) mergers can liberate these tighlty bound stars from their host black holes, resulting in semi-relativistic hypervelocity stars (SHS) that are capable of crossing large swaths of the observable universe and hence can serve as a new cosmological messenger \citep[see also][]{Loeb:2014a}. Predicated on numerical calculations of \citet{Sesana:2010a} and \citet{Iwasawa:2011a} that suggest that BMBHs may be excited to very large eccentricities prior to merger, we argue that many stars originally bound to the less-massive of the two black holes (the secondary) can be ejected in a manner closely resembling the Hills mechanism \citep{Hills:1988a} for the production of HVS. As in the HVS mechanism, these stars can receive a significant speed boost above and beyond their average orbital velocity, occasionally yielding stars with asymptotic velocities $v_{\infty}$ nearing $c$. After several Gyr, these stars evolve off the main sequence and become bright giants that are potentially detectable by future all-sky infrared surveys, and imaging surveys by the next generation of telescopes. We demonstrate that no mechanism aside from eccentric merging BMBHs can accelerate a detectable number of main-sequence (MS) stars with speeds in excess of $\sim 10^{4}$ km~s$^{-1}$, and thus the detection of even a single star moving at a velocity greater than this value would suggest that a significant fraction of BMBH mergers proceed eccentrically. 

This mechanism is schematically depicted in Figure~\ref{fig:diagram}. The merger of two galaxies (panel~1) results in the eventual merger of the nuclear clusters hosting their MBHs (panel~2). As the secondary black hole scatters stars in orbit about the primary, its eccentricity grows quickly to a value of order unity (panel~3). Once the eccentricity has been excited to a large value, the relative binding energy of the secondary to the primary becomes small as compared to the specific binding energy of stars in orbit about either the primary or the secondary, whose eccentricities are on average much lower. As the periapse distance $r_{\rm p,12}$ shrinks, the eccentric Hill radius around the secondary also shrinks, $r_{\rm H,2} = r_{\rm p,12} q_{12}^{\smash{-1/3}}$, where $q_{12} \equiv M_{1}/M_{2}$, resulting in the removal of all stars with apoapses comparable to this distance (panel 4). This process complements the mechanism in which stars that are originally in orbit about the primary occasionally enter the secondary's Hill radius, resulting in their acceleration to similar speeds \citep{Yu:2003a,Holley-Bockelmann:2005a,Levin:2006a,Sesana:2006a}. While we did not perform explicit calculations of this complimentary channel, the mechanism we describe here is qualitatively similar in both the number and distribution of SHS that are produced, which likely would increase the number of SHS by a factor of a few.

\begin{figure}
\centering\includegraphics[width=\linewidth,clip=true]{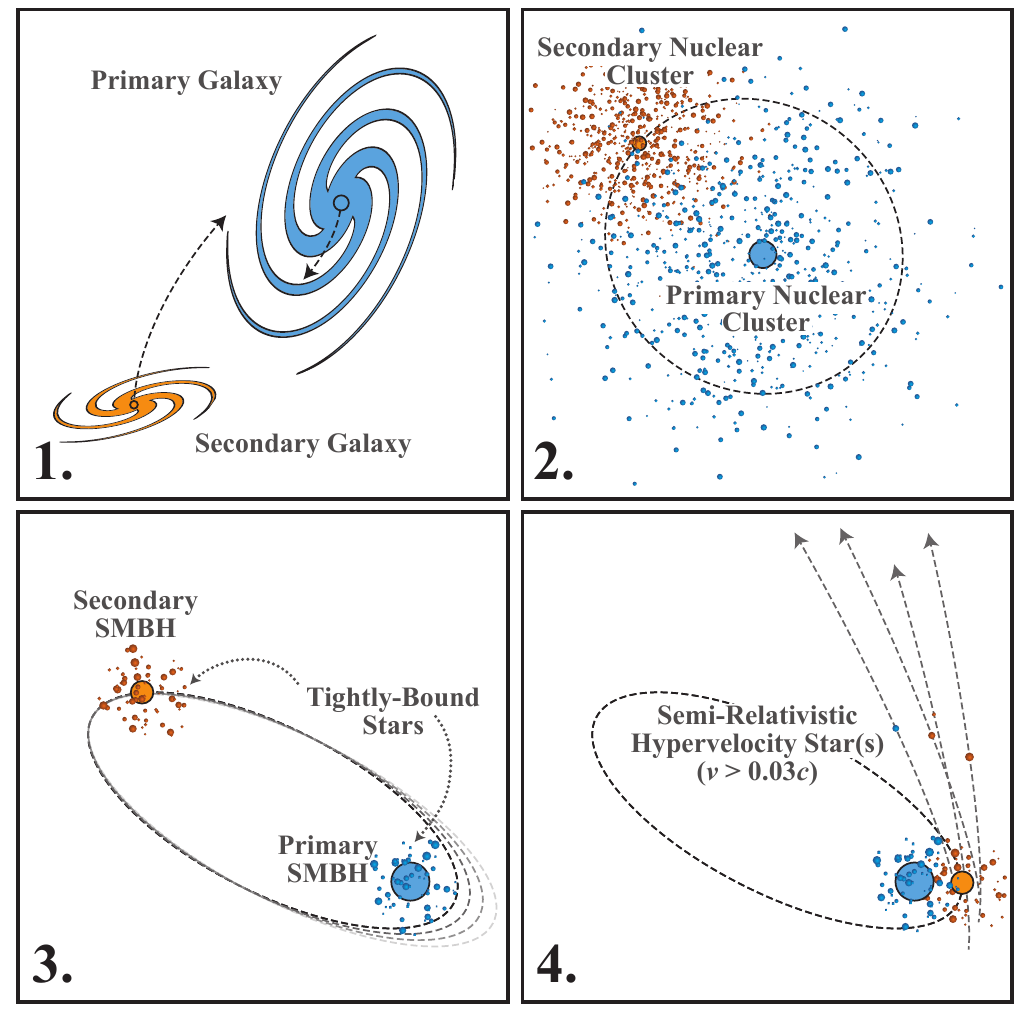}
\caption{Diagram of primary production channel for SHS. 1: Two galaxies with central black holes merge. 2: Dynamical friction brings the two nuclear clusters and their host MBHs together. 3: The eccentricity of the secondary MBH's orbit about the primary is excited by asymmetrical scattering of stars that originally orbited the primary MBH. A tighlty bound cluster of stars remains bound to the secondary. 4: With each passage of the secondary by the primary, a fraction of stars are ejected as SHS.}
\label{fig:diagram}
\end{figure}

\begin{figure*}
\centering\includegraphics[width=\linewidth,clip=true]{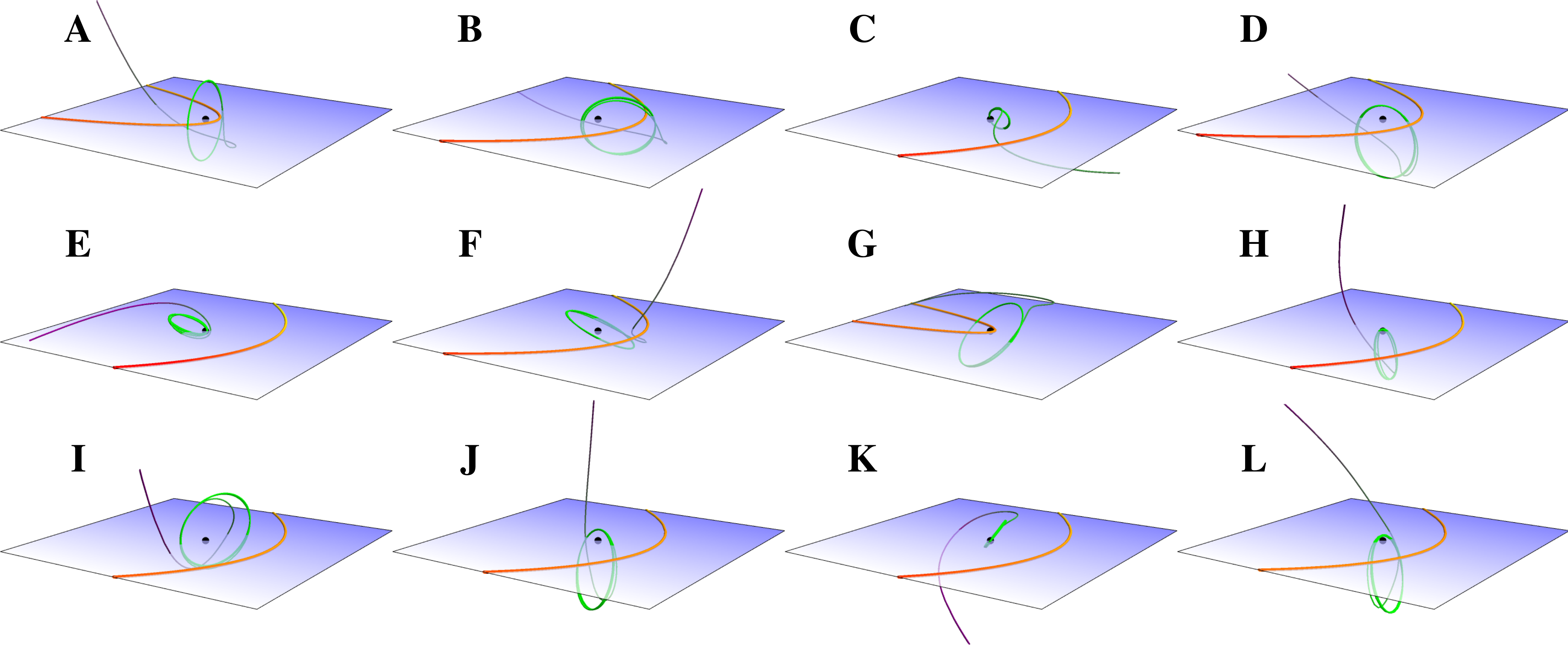}
\caption{Twelve randomly selected trajectories of a stellar ejection resulting from the eccentric passage of the secondary MBH by the more-massive primary MBH, with the semimajor axis of the tertiary restricted to $1~<~\log_{10} a_{23,\min}/r_{\rm IBCO,2}~<~1.25$. Each panel is centered about the secondary MBH's position, with the yellow-red curve showing the trajectory of the primary MBH (yellow early times, red late times), and the green-purple curve showing the trajectory of the ejected star (green early times, purple late times). The initial conditions of each encounter is drawn randomly for each panel, as described in the text. A variety of outcomes is apparent, with the ejected stars leaving in many possible directions relative to the original encounter geometry.}
\label{fig:examples}
\end{figure*}

To quantify the number of SHS produced by merging BMBHs, we perform Monte Carlo three-body scattering experiments in which random combinations of primary, secondary, and tertiary parameters are drawn, and the outcomes are recorded (see Figure~\ref{fig:examples} for example output orbital evolutions). We then measure the outgoing velocity of tertiary objects that escape the system and calculate their subsequent evolution from stellar isochrones in order to determine their detectability via current and future surveys. Additionally, we find that binary star systems (which we term semi-relativistic hypervelocity binaries, or SHB) are also capable of being accelerated by the same mechanism, which may enable additional means for detection and identification of semi-relativistic stars. We find that approximately $10^{6}$ objects moving with velocity greater than $10^{4}$ km s$^{-1}$ exist out to the distance of the Virgo cluster and $\sim 10^{2}$ within a Mpc of the MW. Roughly one SHS should be detectable per 100 square degrees for a survey with limiting K-band magnitude of 24, a number that is too small to guarantee a discovery with current surveys, but will yield tens of detections with a next-generation all-sky infrared survey such as {\it WFIRST}\footnote{\url{http://wfirst.gsfc.nasa.gov}} or {\it Euclid}\footnote{\url{http://www.euclid-ec.org}}.

We find that the number of objects ejected is a strong function of velocity, with the vast majority of SHS that would likely be detected moving with velocity $\lesssim 10^{4}$ km s$^{-1}$. Faster objects would exist near the Milky Way, but these objects are most likely to be low-mass MS dwarfs that would only be detectable with deep imaging by either {\it Hubble Space Telescope}\footnote{\url{http://hubblesite.org}} or {\it James Webb Space Telescope} ({\it JWST})\footnote{\url{http://www.jwst.nasa.gov}}. Most of the detectable stars will be red supergiants that evolved from stars slightly less than a solar mass, but the Doppler shift associated with such large speeds will make these objects appear bluer than a similar star at rest, with color shifts of a few tenths of a magnitude in the bluest {\it JWST} bands. Despite their average distance of a Mpc, their large velocities will give them a proper motion that is potentially detectable with {\it LSST}\footnote{\url{http://www.lsst.org}} over 10 years.

In Section~\ref{sec:speedlimits} we calculate the speed limits for MS stars accelerated via single, double, and triple object mechanisms, and show that only eccentric BMBH mergers can yield MS stars with $v > 10^{4}$ km s$^{-1}$. In Section~\ref{sec:experiments} we describe the setup of our numerical scattering experiments, and in Section~\ref{sec:results} we describe the outcomes of these experiments. In Section~\ref{sec:shb} we demonstrate that binary star systems can also be accelerated by this same mechanism, and present an example ejection. In Section~\ref{sec:observations} we calculate the detectability of luminous SHS with various survey constraints. However, we find that identification may be somewhat more challenging, we discuss the problem of identification and complications of our various model assumptions in Section~\ref{sec:discussion}. alternative means of detection and identification, Lastly, we summarize the scientific value of SHS and SHB if they are detected in Section~\ref{sec:conclusion}.

\section{Speed Limits of MS Stars}\label{sec:speedlimits}
A single star can typically be accelerated to velocities up to its own escape velocity if it can asymmetrically eject a mass comparable to its own mass. Instances of tremendous mass loss usually occur shortly before the birth of a compact object, but mass-loss episodes for stars that do not immediately form compact objects do not result in large bulk velocities for the surviving star. As an example, $\eta$ Carinae, which lost tens of solar masses of ejecta a century ago in an explosive episode, is moving radially toward the Sun at less than 10 km s$^{\smash{-1}}$ \citep{Smith:2004a}.

When two objects interact with one another, the maximum velocity in the system is set by the object with the lowest density, as this object determines the minimum approach distance between the two objects before a collision or tidal disruption occurs. If the two objects are point-like and initially bound to one another, the two objects will remain bound indefinitely, assuming Newtonian dynamics. However, there are three possibilities that enable the ejection of an object at a velocity comparable to the maximum orbital velocity of the system: The tidal break-up of one of the two objects, effectively converting the system into a multi-body encounter \citep{Faber:2005a,Manukian:2013a}, the destruction of one of the two stars via a supernova \citep[runaway stars,][]{Blaauw:1961a}, or the tidal break-up of a binary through interaction with a third object \citep{Hills:1988a,Agnor:2006a,Hoffman:2007a}.

In the first two cases, the outgoing velocity is limited to a fraction of the velocity at pericenter, which is comparable to the escape velocities from the surfaces of the two objects. In the last case, the outgoing velocity can be enhanced by the presence of the third, more massive object \citep{Hills:1988a}. Because of this additional ingredient, this mechanism can produce stars that are significantly faster than what is possible with a system comprised of only one or two objects.

In the remainder of this section we focus upon the combinations of object types with MS stars and examine the maximum velocities that can be produced for a given combination. Label subscripts for each object are assigned in order of descending mass of the constituents, i.e. ``1'' will always refer to the most massive object in the system (the primary), ``2'' to the second-most massive (the secondary), and ``3'' to the least massive (the tertiary). Double subscripts (e.g. ``12,'' ``23'') refer to quantities that jointly apply to both subscripted objects; as an example $a_{12}$ would refer to the semimajor axis of the primary-secondary system. Commas in the subscripts separate groupings, for example $r_{\rm p,1,23}$ would refer to the pericenter distance of the secondary-tertiary system about the primary.

\subsection{The Hills Mechanism}\label{sec:hills}
It was found through numerical experiments by \citet{Sari:2010a} that the maximal velocity for outgoing objects in a Hills encounter ($M_{1} \gg M_{2}$) is
\begin{equation}
v_{\rm kick,\max} = 1.3 v_{23} q_{1,23}^{1/6}\label{eq:vkickmax}
\end{equation}
for the lowest-mass object in the system ($M_{3} < M_{2}$), where $v_{23}$ is the escape velocity from the surface of the secondary object at a separation equal to the sum of the stellar radii,
\begin{equation}
v_{23} = \sqrt{\frac{2 G M_{2}}{R_{2} + R_{3}}}\label{eq:v23},
\end{equation}
and $q_{1,23} \equiv M_{1}/M_{23}$ ($M_{23} \equiv M_{2} + M_{3}$). The highest velocities are guaranteed for objects originating from binaries that are at or near contact, provided that the binary's tidal disruption radius $r_{\rm t,23} \equiv a_{23} q_{1,23}^{\smash{1/3}}$ is greater than the periapse distance $r_{\rm p,1,23}$. If the incoming binary is not a contact binary to begin with, a value similar to that of Equation (\ref{eq:vkickmax}) can be achieved at a distance $\sim 0.1$ that of the tidal disruption radius of the incoming binary \citep{Sari:2010a}, but whether this maximum value is realized depends on the initial phase of the binary's orbit, where optimal phases are those in which the secondary and tertiary objects approach each other as closely as possible at the moment of periapse with the primary, without either colliding or tidally disrupting one another. Because the stars are separated by no less than the sums of their radii, the maximum velocity for wide binaries that are driven to small separations during an encounter is no greater than the disruption of a contact binary. For a maximum kick, the mass ratio $q_{1,23}$ should be as large as possible, but keeping in mind that $v_{23}$ tends to decrease with increasing $q_{1,23}$ for a fixed $M_{1}$. The following sections describe how these two scalings compare to one another for specific object types.

\subsubsection{Stellar-mass Secondaries}\label{sec:stellmasssecond}
The escape velocity from a MS star scales weakly with stellar mass, as $R_\ast \appropto M_\ast^{\smash{0.8}}$ \citep{Tout:1996a}, and thus $v_{\rm esc,\ast} \appropto M_\ast^{0.1}$. Because the kick velocity scales as $q_{1,23}^{\smash{1/6}} \propto M_{2}^{\smash{-1/6}}$ (assuming $M_{2} \gg M_{3}$), there is no advantage to increasing the mass of the secondary, and in fact the maximum kick velocity {\it decreases} slightly, $v_{\rm kick} \appropto M_2^{-0.06}$.

The mass of the black hole that yields the largest velocities is set by finding the largest black hole that will not swallow one (or both) of the stars whole, but still can disrupt the binary. For parabolic encounters, the critical distance at which an incoming object will be swallowed is the ``innermost bound circular orbit'' $r_{\rm IBCO}$ \citep{Bardeen:1972a}, and for non-spinning black holes this distance is twice the Schwarzschild radius, $r_{\rm IBCO} = 4 r_{\rm g} = 4 G M / c^{2}$. In principle, the binary separation $a_{23}$ can be arbitrarily large and still experience a large kick, provided that the two stars approach one another closely at periapse, but the probability of this occurring for a given passage is progressively smaller for increasing initial separations.
%Binaries of separations greater than
%\begin{equation}
%a_{23,\max} = \frac{G M_{23}}{\sigma^2},\label{eq:amax}
%\end{equation}
%where $\sigma$ is the velocity dispersion at the sphere of influence of the central cluster, are unlikely to survive random encounters with other stars, and thus this sets a limit on the maximum separation. With this in mind, increasing $M_{23}$ can help achieve larger kick velocities than what would otherwise be possible as these binaries can more easily survive in the nuclear clusters surrounding massive black holes.

To maximize Equation (\ref{eq:vkickmax}), we want to determine the most massive primary that will not swallow a given binary whole to maximize the $q_{1,23}$ ratio, and thus we set the periapse distance $r_{\rm p,23} = r_{\rm IBCO,1}$. If we presume that the binary is a contact binary, $a_{23} = (R_{2} + R_{3})/2$, and assuming $R_{\ast} = R_{\odot} (M_{\ast}/M_{\odot})^{\smash{0.8}}$, we find the primary mass $M_{1}$ that leads to the largest outgoing velocity given a stellar mass,
\begin{equation}
M_{1,\max} = 4 \times 10^{7} \left(\frac{M_{2}}{M_{\odot}}\right)^{0.7} M_{\odot}.
\end{equation}
Combing this expression with Equation (\ref{eq:vkickmax}), we find the maximal velocity possible for a star of a given mass, independent of $M_{1}$,
\begin{equation}
v_{\rm kick,\max} = \textrm{11,000} \left(\frac{M_2}{M_{\odot}}\right)^{0.05} \textrm{km~s$^{-1}$}\label{eq:vkickmax2},
\end{equation}
which scales extremely weakly with $M_{2}$. Any star moving faster than this velocity is very unlikely to be produced by the standard Hills mechanism in which both objects are MS stars, as it would require that a binary with a wide separation (and thus $v_{\rm circ} < \sigma$) passes by the central black hole, which is only possible if the binary is set onto a plunging orbit from beyond the black hole's sphere of influence. This limit is consistent with numerical experiments \citep{Ginsburg:2012a}.

By changing the secondary to a degenerate object, such as a WD or a NS, $M_{2}$ is limited to the Chandrasekhar mass $M_{\rm Ch} \simeq 1.4 M_{\odot}$. This means that either the tertiary is selected to be somewhat less massive than the secondary, or that the compact object becomes the tertiary, limiting the ejection velocity of the (now heavier) MS companion to $0.9 v_{\rm esc} q_{1,23}^{\smash{1/6}}$ \citep{Sari:2010a}. In either case, the maximum kick velocity is smaller than what is possible with two MS stars, whose masses can exceed $M_{\rm Ch}$, although only slightly given the weak $M_{2}$ dependence of Equation (\ref{eq:vkickmax2}).

The only option for secondaries with masses larger than that of the most massive stars ($\sim 300 M_{\odot}$) are black holes. If the secondary is a stellar-mass black hole (SBH), $M_{2}$ is not limited by $M_{\rm Ch}$, and the denominator of Equation (\ref{eq:v23}) ceases to depend on $R_{2}$ as $R_{3} \gg R_{2} = r_{\rm IBCO,2}$ for $M_{2} \lesssim 10^{5} M_{\odot}$. However, as the secondary becomes more massive, and its separation is only determined by the size of the tertiary, the tertiary can be tidally disrupted by the secondary, and this tidal disruption radius $r_{\rm t, 3} = R_{3} q_{23}^{\smash{1/3}}$ sets the minimum separation distance. Incorporating these changes, the kick velocity scales with $M_{2}$ to a higher power, although still rather weakly,
\begin{equation}
v_{\rm kick,\max} = \textrm{20,000} \left(\frac{M_2}{10 M_{\odot}}\right)^{1/6} \left(\frac{M_3}{M_{\odot}}\right)^{-0.12} \textrm{km~s$^{-1}$}\label{eq:vkickmax3},
\end{equation}
where above we have scaled $M_{2}$ to a $10 M_{\odot}$, the mass of a typical SBH. The expression above applies for $M_{2} > M_{3}$ and $M_{3} > 0.1 M_{\odot}$. The most-massive stellar mass black hole known may have a mass as large as $33 M_{\odot}$ \citep{Silverman:2008a}, yielding a $v_{\rm kick,\max} = $~24,000~km~s$^{-1} = 0.08 c$ for a solar mass star.

\subsubsection{Massive Secondaries}
Black hole masses are known to range from a few to tens of billions of solar masses, with a paucity of black holes known to have masses between 100 and $10^{5} M_{\odot}$. If this deficit is real, then it suggests that stars launched via the Hills mechanism where the secondary is a black hole have a bimodal distribution of velocities. As we showed in the previous section, the limit for MS binaries is $\sim 0.05 c$, and this limit arises from the fact that MS stars of increasing mass have increasing size, and thus collide rather than resulting in an ejection. Black holes do not have this limitation, so in principle the maximum velocity possible for a non-spinning black hole could be as large as the velocity at the Schwarzschild radius of the secondary $r_{\rm IBCO,2}$, the speed of light $c$.

Equation (\ref{eq:vkickmax3}) estimates the maximum kick velocity possible so long as $r_{\rm IBCO,2}$ is small compared to $R_{3}$, but the picture is of course complicated by general relativity, which can cause particles on orbits of finite energy to inspiral exterior to $r_{\rm IBCO,2}$, depending on the black hole's spin. As a hypervelocity ejection is typically the result of the tertiary being placed on a near-radial orbit about the secondary by the primary, it approaches the secondary on a parabolic orbit, which would inspiral into the secondary at its IBCO at $r_{\rm IBGO,2}$ for a non-spinning black hole. As a result, the maximum possible velocity is set at this distance rather than the Schwarzschild radius $r_{\rm Sch,2} = 2 r_{\rm g,2}$. This suggests that the speed limit for SHS is $c/2$ (including the relativistic boost factor $\gamma$), and possibly more for spinning black holes, which we do not consider here.

This maximum speed is a factor $\sim 10$ times larger than the speed limit for MS binaries, and 5 times larger than the limit for MS-SBH binaries. If the spectrum of SHS velocities were similar to that of HVS, in which the typical velocities are a few thousand km~s$^{-1}$, it would suggest that the average SHS's speed could be a few $10^{4}$ km~s$^{-1}$. However, two important factors influence the SHS velocity distribution: the secondary's binding energy to the primary, which determines the depth of the potential well that any ejected tertiary would need to climb out of, and the distribution of orbits about the secondary, which is quite different from the distribution of orbits of stellar binaries.

\subsection{Importance of Secondary's Orbital Energy}\label{sec:orbenergy}
For the traditional HVS mechanism, the incoming binaries are deposited into the loss cone via kicks they receive typically near apoapse, with the majority of these binaries originating from the primary's sphere of influence \citep{Lightman:1977a}. At this distance, the binding energy to the black hole is initially small ($\sim \sigma^{2}$), and thus the final velocities of ejected stars are only reduced by a small amount \citep{Kobayashi:2012a}.

For the SHS mechanism, the secondary's orbital energy depends entirely on how the two black hole clusters merge. The traditional picture has been that dynamical friction drags the secondary black hole and its surrounding stars into the outskirts of the primary's nuclear cluster. At this point, stars are scattered by the secondary with no preferred direction, resulting in an inspiral in which the secondary's orbit remains approximately circular at all times, although eccentricity may be excited for shallow stellar density profiles \citep{Antonini:2012a}. This means that stars that are tidally removed from the secondary by the primary will only leave with a velocity comparable to the local velocity dispersion, and will not escape the primary's gravity.

However, numerical $N$-body results \citep{Baumgardt:2006b,Iwasawa:2011a,Khan:2012a} suggest that the secondary's orbit does not remain circular during its inspiral owing to a preferential ejection of stars orbiting the primary in the same direction as the secondary. They suggest that the orbit can become extremely eccentric, with $(1 - e) \sim M_{2}/M_{\ast} = 10^{-8}$ (ignoring general relativistic effects) for $M_{2} = 10^{8} M_{\odot}$. \citet{Iwasawa:2011a} suggests that the only mechanism that prevents a total plunge ($e = 1$) is the random noise introduced by the discrete nature of the scattered stars, which suggests that more massive secondaries could result in even more extreme maximum eccentricities. If such behavior occurs in nature, the specific orbital energy of the secondary black hole is comparable to that of incoming stellar binaries in the HVS scenario, and thus the velocities of ejected stars will not be reduced much by their initial orbital energy.

\section{Rates}

\subsection{Scattering of Stars from the Primary Cluster}

When a cluster contains a single MBH, there are two primary mechanisms for ejecting stars at great velocities. The first mechanism is the Hills mechanism \citep{Hills:1988a}, described in Section~\ref{sec:hills}. The second is the scattering of a star by a random passer-by star or compact object \citep{Binney:1998a,OLeary:2008a}. Although the initial configuration pre-encounter is different in the two cases, the setup of the system at the time of scattering is ultimately the same: Two stars that lie a distance from one another that is less than their mutual Hill radii. This similarity can be seen by comparing expressions presented for the two mechanisms in \citet{Yu:2003a}; setting the distance of closest approach equal to the binary separation results in a velocity different of $\sqrt{2}$, the difference in orbital velocity between a circular and parabolic orbit. The scaling of $v$ with the masses of the two stars is identical for the two mechanisms.

The literature concerning HVS produced via the mergers of MBH has focused on the stars originally bound to the primary \citep{Quinlan:1996a}, more specifically, the scenario of a binary MBH in our own galactic center was considered for producing HVS \citep{Yu:2003a,Sesana:2007b}. These papers focused on the analog of the scattering by a random passer-by, except in this case the passer-by is itself a MBH. As above for stars, the Hills mechanism and scattering by a random passer-by are identical modulo a factor $\sqrt{2}$ for given secondary-tertiary separation. In both cases, the total mass of stars liberated by the process is on order the mass of the secondary, in the passer-by case this has to do with the fraction volume occupied by the secondary's Hill radius, $V_{2}/V_{1} = (r_{\rm H,2}/r_{1,2})^{3} = M_{1}/q_{1,2} = M_{2}$, whereas in the Hills case the mass liberated is simply the mass of stars originally bound to the secondary $M_{2}$.

\subsection{Removal of Secondary's Cluster}
In the initial stages of a merger, the secondary retains a group of stars within its Hill radius, whose distribution likely resembles the original distribution of stars. Prior to merger, the radial distribution of these stars likely resembles $n(r) \propto r^{-7/4}$ \citep{Bahcall:1976a} exterior to the distance where the orbital velocity $v_{\rm orb} = \sqrt{G M_{2} / r}$ is less than the escape velocities from typical stars, $v_{\ast} = \sqrt{G M_{\ast} / R_{\ast}}$. If the black hole is surrounded by a population of stellar mass black holes, this distribution can continue a factor of a few more in $r$ before terminating, as stars would be capable of relaxing to higher binding energies without colliding \citep{OLeary:2008a}. In such systems, $N(a,e) \propto a^{1/4} e$ \citep{Merritt:2013a}. However, if we consider a scenario where the cluster orbits within a larger cluster on an eccentric orbit, the boundary condition of the secondary's cluster is complicated. Not only does the region within which the secondary dominates the dynamics change in size, the conditions at its boundary change with distance from the central black hole.

If we assume that the cusp of stars around the primary follows a power-law distribution with $r$, $\rho \propto r^{-\alpha}$, where $\alpha < 2$, then the stalling radius for the secondary is given by the distance within which a mass $2 M_{2}$ of stars is contained \citep{Matsubayashi:2007a},
\begin{equation}
a_{\rm s} = a_{\rm h,1} q_{12}^{\frac{1}{\alpha-3}},
\label{eq:astall}
\end{equation}
where $a_{\rm h,1} \equiv G M_{1}/\sigma_{1}^{2}$ is the radius of the sphere of influence of the primary. At this distance, the stellar density is
\begin{equation}
\rho(a_{\rm s}) = \frac{3 - \alpha}{2\pi} q_{12}^{\frac{\alpha}{3 - \alpha}} \frac{M_{1}}{a_{\rm h,1}^{3}}.
\label{eq:rhostall}
\end{equation}

At the primary's sphere of influence, the velocity dispersion is \citep{Kormendy:2013a}
\begin{equation}
\sigma_{1} = 75 \left(\frac{M_{1}}{4 \times 10^{6} M_{\odot}}\right)^{0.25} \textrm{km}\label{eq:sigma},
\end{equation}
and so the Hill radius of the secondary at this location is
\begin{equation}
r_{\rm H} = 3^{-1/3} q_{12}^{1/3} \frac{G M_{1}}{\sigma_{1}^{2}}.
\end{equation}

While the secondary's eccentricity grows, the Hill radius shrinks as the secondary's periapse comes closer to the primary. This results in the stripping of the outermost stars in orbit about the secondary. As the eccentric growth rate is slow compared to the orbital period of stars around the secondary, stars are almost always removed when their apoapse is larger than secondary's eccentric Hill sphere at periapse,
\begin{equation}
r_{\rm H,e} = (1 - e_{12}) r_{\rm H}.
\end{equation}
This means that stars with the smallest separations from the secondary are removed in the final phases of the secondary's inspiral. These stars are the ones that potentially have the largest ejection velocities, so long as the secondary's orbit remains eccentric during the inspiral. Eventually, every single star that was once bound to the secondary will experience one of the following outcomes: become bound to the primary, be swallowed or tidally disrupted by one of the two black holes, or become unbound to both the primary and the secondary. This last possibility can potentially produce SHS.

\subsubsection{Justification for Single-scattering Approximation}\label{sec:singlescat}
The secondary black hole's eccentricity increases slowly as a function of time, meaning that the secondary-tertiary system gradually becomes more and more prone to perturbation from the primary. In principle, the tertiary can be lost well before $a_{23} > r_{{\rm H},23}$, meaning that the maximum kick velocity may be limited \citep{Sari:2010a}. However, if the evolution of the secondary's eccentricity is rapid, the periapse distance may change significantly from orbit to orbit, and the periapse distance $r_{\rm p,2}$ for a particular three-body encounter should be drawn randomly between zero and $r_{{\rm H},e}$ \citep[i.e. ``pinhole'' scattering,][]{Lightman:1977a}. In such a situation, the first strong interaction for a given tertiary may be at a periapse distance that is equal to or even significantly deeper than the tidal disruption radius.

\citet{Iwasawa:2011a} presented an expression for the eccentricity growth timescale $T_{e}$ as a function of the primary-secondary binary parameters, which we reproduce here using our notational conventions,
\begin{align}
T_{e} &\sim 1.4 \times 10^{8} \left(\frac{M_{1}}{10^{10} M_{\odot}}\right)^{3/2}\left(\frac{M_{2}}{10^{8} M_{\odot}}\right)^{-1}\nonumber\\
&\times \left(\frac{a}{10\;{\rm pc}}\right)^{3/2}\left(\frac{\rho}{100 M_{\odot}\;{\rm pc}^{-3}}\right)^{-1} {\rm yr}.
\label{eq:te}
\end{align}
The main uncertainty in this expression the values of $a$ and $\rho$. When two central black holes merge, their evolution is initially governed by dynamical friction, which causes the lighter of the two black holes to sink rapidly toward the heavier \citep{Dotti:2012a}. During this phase, the orbit may acquire moderate eccentricities for shallow density profiles, but often remains circular \citep{Antonini:2012a}. This phase ceases once the secondary black hole reaches the stalling radius $a_{\rm s}$ (Equation (\ref{eq:astall})), at which point $a$ ceases to evolve for isotropic stellar distributions. At this point, the eccentricity of the secondary begins to grow, and thus the appropriate values of $a$ and $\rho$ are determined at a distance $a_{\rm s}$ from the primary.

Setting $a = a_{\rm s}$ (Equation \ref{eq:astall}) and $\rho = \rho(a_{\rm s})$ (Equation \ref{eq:rhostall}) in Equation (\ref{eq:te}) and taking the ratio of this timescale to the secondary's orbital period $P_{2}$ at $a_{\rm s}$, we find
\begin{equation}
%\frac{T_{e}}{P_{2}} = \frac{2.4 \times 10^{-4}}{3 - \alpha} q_{12}^{2}.
\frac{T_{e}}{P_{2}}=\frac{0.336}{(3-\alpha)}\left(\frac{M_1}{10^{9} M_{\odot}}\right)^{3/2}q_{12}^{\frac{2\alpha-3}{\alpha-3}}
\end{equation}
The orbit of the secondary can be considered ``plunging'' if $T_{e}$ is comparable to $P_{2}$. Setting $T_{e}/P_{2} = 1$ and solving for $q_{12}$ we find the maximum mass ratio that will have rapid eccentricity evolution $q_{\rm plunge}$,
\begin{equation}
q_{\rm plunge} = 65 \sqrt{3-\alpha}\label{eq:qplunge},
\end{equation}
which is equal to 73 for $\alpha = 7/4$. However, it has been shown that near-equal mass MBHs ($q_{12} \lesssim 3$) do not show strong eccentricity growth \citep{Sesana:2010a,Wang:2014b}. As $T_{e}/P$ increases with increasing $q_{12}$, this suggests that there must be a critical $q_{12}$ value for which eccentricity excitation is the most effective between these two extremes, however any mass ratio in between these two values should have $P_{2} \lesssim T_{e}$, and thus a single-scattering approximation in which $r_{\rm p,2}$ is randomly drawn is appropriate for these encounters.

\subsection{Similarities Between Primary and Secondary SHS}\label{sec:similarity}
While the two scenarios described above are somewhat different in their initial setup, the conditions required to produce an SHS are remarkably similar. For stars that are originally bound to the primary (primary SHS), most of the encounters with the secondary will occur at a distance from the secondary comparable to the secondary's Hill radius. Likewise, stars that are original bound to the secondary (secondary SHS) are removed from the secondary when the Hill radius shrinks to a size comparable to their distance from the secondary. As we show in Section~\ref{sec:results}, the resulting distribution of secondary SHS is very similar to that found for primary SHS. Additionally, the total number of primary and secondary SHS are approximately the same, with the total mass ejected being $\simeq M_{2}$ in both cases.

If the eccentricity of the secondary is small, primary SHS will only acquire an energy comparable to the secondary's circular velocity about the primary. The same is true for secondary SHS; stars that are removed from a secondary on a near-circular orbit will also have velocities comparable to the secondary's circular velocity. The high-velocity tail is only accessible for primary SHS when the secondary has a significantly eccentric orbit \citep{Sesana:2006a}, and as is described in Section~\ref{sec:orbenergy}, the same is true for secondary SHS. Because these two mechanisms are the only channels that produce such high-velocity stars, and the highest velocities are only accessible in eccentric mergers, the discovery of SHS would suggest that BMBH indeed merge eccentrically.

\section{BMBH Scattering Experiments}\label{sec:experiments}
To determine the rate of stellar ejections, we perform Monte Carlo scattering experiments using the same orbit integration routines in {\tt Mathematica} (version 9.0) of \citet{Manukian:2013a} used to investigate the production of ``turbovelocity'' stars. While this approach is only modestly scalable to multiple processors on a single machine, it has the advantage of completely controllable numerical errors, at the expense of increased computational cost. For this paper we perform our scattering experiments with quadruple floating-point precision ($\simeq 32$ significant decimal digits), and restrict the maximum relative error in the orbital energy $E$ and the vectorial angular momenta $J_{x}$, $J_{y}$, and $J_{z}$ to $\simeq 10^{-14}$. This extreme precision enables us to evaluate the back-reaction on the black holes imparted by the ejection of the SHS, which is not possible when the primary-secondary system is assumed to have a fixed orbit. For the special case of the ejection of binary stars, described in Section~\ref{sec:shb}, we employ octuple floating-point precision ($\simeq 64$ digits).

We are ultimately interested in the typical black hole and stellar properties that produce SHS within a given velocity range. As the mechanisms we describe here hinge upon MBH mergers at the centers of galaxies, we need to determine the merger rates of the galaxies themselves, and relate the properties of those merging galaxies to the black holes that they host. The parameter space to be explored is quite large; known MBHs range over five orders of magnitude in mass, and the stars that orbit them possess a range of masses and orbital parameters. Additionally, the rate of BMBH mergers is very poorly quantified, especially for those black holes that are separated by less than $\sim$10 pc (a typical value for $a_{\rm stall}$), which are difficult to resolve even in the radio \citep{Burke-Spolaor:2011b}. While the rates of merging MBHs is highly uncertain, the rates of dark matter halos are very well known through via results of dark matter-only simulations such as the Millennium simulations \citep{Boylan-Kolchin:2009a,Genel:2010a,Fakhouri:2010a}. We use these results as a starting point for determining the rate of MBH mergers.

We use a standard Sheth-Tormen distribution for determining dark matter halo masses ${\cal M}_{\rm h}$, calculated through the online tool {\tt HMFcalc}\footnote{\url{http://hmf.icrar.org}} \citep{Murray:2013a} using the default parameters, and taking $z = 0$. We draw $4 \times 10^{4}$ primaries from this distribution, and then calculate the stellar mass ${\cal M}_{\ast}$ from ${\cal M}_{\rm h}$ via Equation (22) of \citet{Moster:2010a},
\begin{equation}
{\cal M}_{\ast} = {\cal M}_{\ast,0} \frac{\left({\cal M}_{\rm h}/{\cal M}_{1}\right)^{\gamma_{1}}}{\left[1+\left({\cal M}_{\rm h}/{\cal M}_{1}\right)^{\beta}\right]^{\left(\gamma_{1}-\gamma_{2}\right)/\beta}}
\end{equation}
where we have changed the variable notation slightly to match our conventions, and the constants are as defined in that paper: $\log {\cal M}_{\ast,0} = 10.864$, $\log {\cal M}_{1} = 10.456$, $\gamma_{1} = 7.17$, $\gamma_{2} = 0.201$, $\beta = 0.557$. When drawing a stellar mass we randomly add 0.15 dex of variance to the returned value, the intrinsic scatter in the relation as noted by \citeauthor{Moster:2010a}

Using the results of \citet{Bluck:2014a} (see their Figure~2), we randomly draw a bulge-to-total stellar mass ratio ${\cal M}_{\rm bulge}/{\cal M}_{\ast}$ for each galaxy and multiply the total stellar mass by this value to obtain the bulge mass ${\cal M}_{\rm bulge}$. Finally, we use the bulge to black hole mass relation determined by \citet{McConnell:2013a} to obtain $M_{1}$, the mass of the primary black hole in the system,
\begin{equation}
\log_{10} M_{1} = 8.46 + 1.05 \log_{10} \left[\frac{{\cal M}_{\rm bulge}}{10^{11} M_{\odot}}\right],
\end{equation}
adding 0.34 dex of scatter to the returned value, as reported in that work.

With the primary's black hole mass now determined, we now refer to the results of \citet{Fakhouri:2010a} in which they present the expected number of mergers for a halo of a given mass ${\cal M}_{\rm h,1}$ at a given redshift $z$ with secondaries with a mass ratio $\xi_{12} = {\cal M}_{\rm h,2}/{\cal M}_{\rm h,1}$. We note with this definition that we are implicitly assuming the merger rate is defined by the mass of the descendant halo rather than the mass of the progenitor halo, this introduces a slight bias toward higher masses for the highest-mass halos \citep{Genel:2009a}.

For each primary host halo, we integrate Equation (1) of \citet{Fakhouri:2010a},
\begin{equation}
\frac{{\rm d}N_{\rm m}}{{\rm d}\xi {\rm d}z} = A \left(\frac{{\cal M}_{1}(z)}{10^{12} M_{\odot}}\right)^{\alpha} \xi^{\beta} \exp \left[\left(\frac{\xi_{12}}{\tilde{\xi}}\right)^{\gamma}\right] \left(1 + z\right)^{\eta},\label{eq:mergerrate}
\end{equation}
where $A$, $\alpha$, $\beta$, $\gamma$, $\tilde{\xi}$ and $\eta$ are constants defined in that paper, from $z = 0$ to $z = 10$ and $\xi_{12} = 10^{-3}/{\cal M}_{1}$ to $\xi = 1$ (see Section~\ref{sec:singlescat}) to obtain the average number of mergers for a halo of that mass. The mass of a halo as a function of redshift ${\cal M}(z)$ is determined by selecting a halo mass at $z = 0$ and integrating the equation for the median growth rate of a halo,
\begin{align}
\frac{d{\cal M}_{1}}{dz} &= 25.3 M_{\odot}~{\rm yr}^{-1} \left(\frac{{\cal M}_{1}(z)}{10^{12} M_{\odot}}\right)^{1.1}\nonumber\\
&\times\left(1 + 1.65 z\right) \sqrt{\Omega_{\rm m} \left(1 + z\right)^{3}+\Omega_{\Lambda}}\label{eq:mediangrowth},
\end{align}
the second line of Equation (2) of \citeauthor{Fakhouri:2010a}, where $\Omega_{\rm m}$ are $\Omega_{\rm \Lambda}$ are the standard cosmological parameters and are taken to be $\Omega_{\rm m} = 0.27$ and $\Omega_{\Lambda} = 0.73$. The number of mergers $N_{\rm m}$ is then used as an input to a Poisson distribution, from which we draw the number of mergers experienced by that halo $N_{\rm merger}$. Then, we draw $N_{\rm merger}$ realizations of $\xi$ over the same range defined above using Equation (\ref{eq:mergerrate}), where we assume that $\xi$ has no dependence on $z$ or ${\cal M}$. This enables a calculation of the halo mass of the secondary ${\cal M}_{2} = \xi_{12}{\cal M}_{1}$, from which we calculate the secondary's black hole mass $M_{2}$ using the same procedure used above to determine $M_{1}$.

Because mergers of near-equal mass will likely not result in large eccentricities, and because secondaries with too small of a mass will not have rapid eccentricity evolution (see Section~\ref{sec:singlescat}), we eliminate all mergers for which $q_{12} < 3$ or $q_{12} > q_{\rm plunge}$. This restricts our calculation to systems in which the secondary's eccentricity grows to large values on a short timescale and our single-scattering approximation is applicable. As can be seen in Figure~\ref{fig:qplunge}, this reduces the total number of SHS produced to about one third the amount that would be produced if all SBH mergers were eccentric and plunging.

\begin{figure}
\centering\includegraphics[width=\linewidth,clip=true]{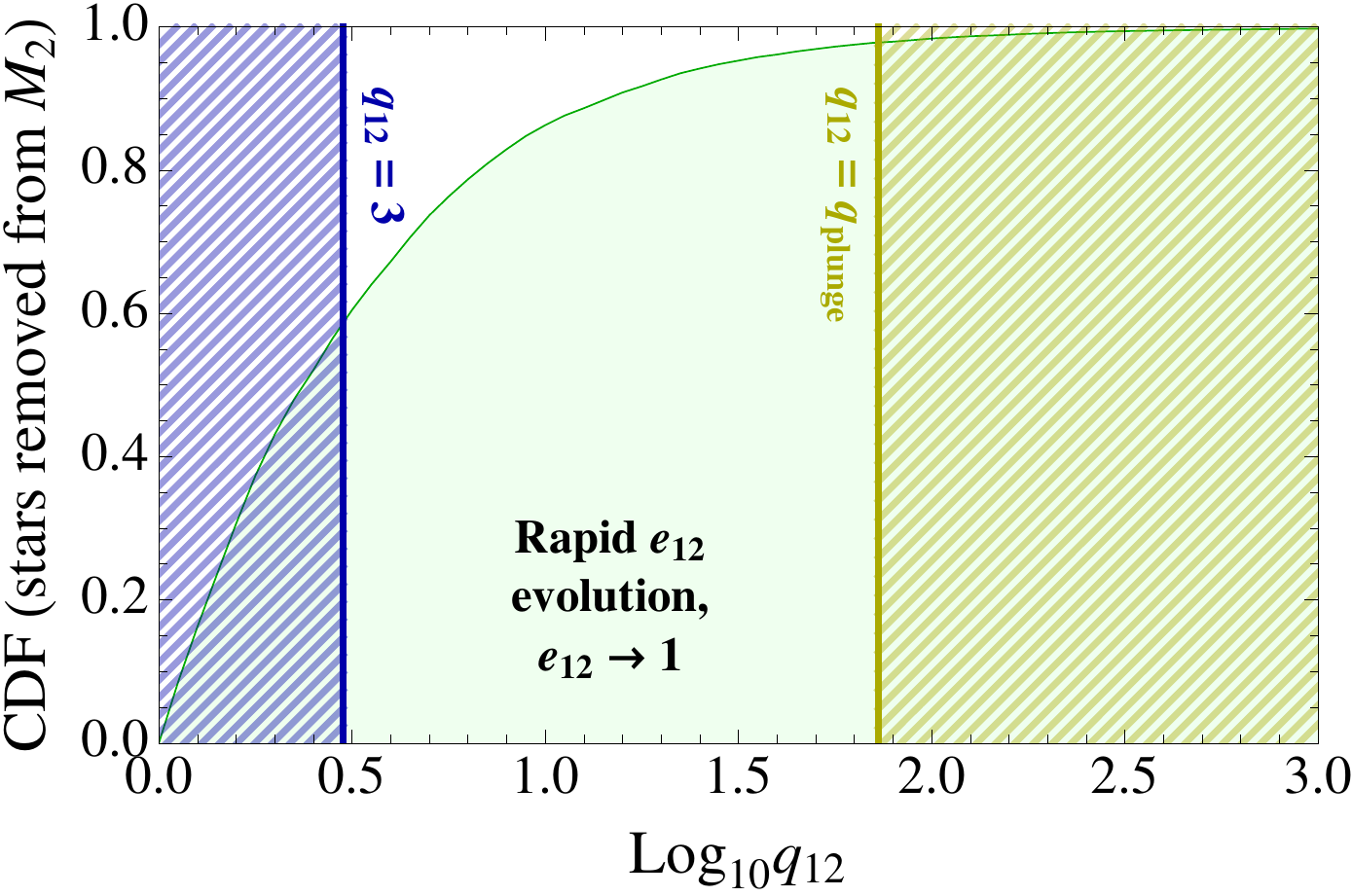}
\caption{Cumulative distribution function of hypervelocity stars produced from mergers of a given $q_{12}$. Below a critical mass ratio $q_{\rm plunge}$ (Equation~(\ref{eq:qplunge})), the cross-section of the secondary is small, reducing its dynamical drag and rate of eccentricity excitation \citep{Iwasawa:2011a}. Above $q_{\rm plunge}$, the secondary's eccentricity evolves on a timescale comparable to an orbital period. For $q_{12} \lesssim 3$ \citep{Sesana:2010a}, eccentricity excitation ceases to be effective, as both black holes begin to orbit a common center of mass.}
\label{fig:qplunge}
\end{figure}

With sample of merger events now defined, three-body encounters are drawn from the sample of mergers with the probability of a draw being proportional to $M_{2}$, as the total number of stars within the secondary's sphere of influence scales directly with its mass, $M_{\ast,2} = 2 M_{2}$. This implies that the most massive mergers can potentially produce the most SHS, for the simple reason that more massive black holes are orbited by more stars. The properties of the tertiary (the star in the system) are assumed to not depend on either black hole; its mass $M_{3}$ is drawn from a Kroupa distribution \citep{Kroupa:2001a}, and its orbital parameters relative to the secondary are drawn presuming a thermal ($P(e) \sim e$), isotropic distribution of orbits about the secondary.

A system of identically massive gravitating bodies in orbit about a dominant central object will relax via two-body interactions to a radial distribution $n(r) \propto r^{-\gamma}$, where $\gamma~=~7/4$, corresponding to a distribution in semimajor axis of $P(a)~\propto~a^{2-\gamma}~\propto~a^{1/4}$ \citep{Bahcall:1976a}. Two-body relaxation becomes ineffective interior to a distance $a_{\rm relax}$ in which the timescale for star-star collisions is shorter than the relaxation timescale. If a population of stellar mass black holes are present in the cores of nuclear clusters, $a_{\rm relax}$ for a solar-type star is determined by where the escape velocity from the stellar mass black hole is equal to the local Kepler velocity about the secondary, $a_{\rm relax} = (M_{2} / M_{\rm bh}) R_{\odot}$, where we assume $M_{\rm bh} = 15 M_{\odot}$ \citep{OLeary:2008a}. The distribution of low-mass MS stars is poorly known interior to $a_{\rm relax}$, as we are only able to resolve individual MS stars within $a_{\rm relax}$ for our own galactic center. For these stars, for which only high-mass MS stars ($M_{\ast} \gtrsim 4 M_{\odot})$ with short lifetimes are detectable \citep{Ghez:2005a}, the distribution with distance from the black hole $r$ is very shallow, $P(r) \propto r^{-0.5}$, implying $P(a) \propto a^{1.5}$. However, the surface brightness distribution of the unresolved stars in the galactic center, which is thought to be dominated by K-dwarf stars, is slightly steeper, $P(r) \propto r^{-1}$ \citep{Yusef-Zadeh:2012a}, implying $P(a) \propto a$. Because it turns out that most of the observed SHS are giants that evolved from relatively low-mass MS stars, we use the power-law implied by the unresolved population, $P(a_{23}) \propto a_{23}$, for $a_{23} < a_{\rm relax}$.

Given these functional forms, the tertiary's semimajor axis distribution is defined as a broken power-law, with a cutoff corresponding to the maximum of the tidal radius of the tertiary and the Schwarzschild radius of the secondary $a_{\rm dest} \equiv \max$($r_{\rm IBCO,2}$, $q_{23}^{\smash{1/3}} R_{3}$), and a cutoff excluding objects in which the gravitational wave merger timescale $\tau_{\rm GW,23}$ of the tertiary about the secondary \citep[see][]{Peters:1964a} is less than the orbital period of the secondary about the primary $P_{12}$,
\begin{eqnarray}
P(a_{23}) \propto
\begin{cases}
a_{23}^{1/4} &: a_{23} > a_{\rm relax}\\
a_{23} &: a_{23} < a_{\rm relax}\\
\end{cases},\label{eq:adist}\\
P(a_{23}) = 0 {\rm\;if\;} a < a_{\rm dest} \lor \tau_{\rm GW,23} < P_{12}.\nonumber
\end{eqnarray}

Each simulation is started at a time $t_{0} = t_{\rm p,12} - 10 P_{23}$, where $t_{\rm p,12}$ is the time of the secondary's periapse about the primary and $P_{23}$ is the orbital period of the tertiary about the secondary; as the encounters are all close to parabolic this timespan is sufficiently long to ensure that the secondary-tertiary system is unperturbed by the primary at $t~=~0$. The choice of an integer number of orbital periods also ensures that effects of the phase of the orbit, defined by the initial mean anomaly $M_{0}$, can be directly related to the experiment outcomes.

Rather than draw tertiaries directly from Equation (\ref{eq:adist}), which would result in a distribution in which low-energy stars are better sampled than high-energy stars, we perform a number of independent samples within small bins of $a_{23}$, keeping in mind that the outcomes of each experiment needs to be normalized later by integrating Equation (\ref{eq:adist}) over the range of sampled $a_{23}$. We select a bin size of 0.25 dex in $\log_{10} \tilde{a}$, where we define $\tilde{a} \equiv a_{23} / r_{\rm IBCO,2}$, the ratio of the tertiary's semimajor axis to the IBCO of the secondary, with our bins spanning $1~\leq~\tilde{a}~\leq~10^{7}$. 4096 systems are then drawn for each bin; with 28 bins this means that our results are derived from 114,688 independent three-body experiments.

Lastly, we must select a proper periapse distance for each encounter. For circular orbits, the region of stability is well-defined by the Jacobi constant, but this expression ceases to be constant when generalized to an elliptical orbit. Numerical experiments have shown that triple systems will eventually lose one of their components at periapse distances that are significantly greater than the tidal disruption radius of the inner binary $r_{\rm H,23}/a_{23} = 3^{-1/3} q_{12}^{\smash{1/3}}$ \citep{Mardling:1999a,Mardling:2001a},
\begin{align}
\frac{r_{\rm p,12}}{a_{23}} &< 2.8 \left[\left(1 + q_{12}\right) \frac{1 + e_{12}}{\left(1 - e_{12}\right)^{1/2}}\right]^{2/5} \left(1 - \frac{0.3 i_{23}}{180}\right)\label{eq:rpcrit}\\
&\equiv \frac{r_{\rm p,12,crit}}{a_{23}},\nonumber
\end{align}
where $i_{23}$ is the inclination of the orbital plane of the secondary-tertiary system relative to the orbital plane of the primary-secondary system, in degrees. Because stability is guaranteed for systems with greater $r_{\rm p,12}$, this expression sets the maximum possible $r_{\rm p,12}$ for which ejection is possible, and thus we only draw $r_{\rm p,12}$ values that are less than this limit. As the secondary is in a plunging orbit in which its orbital angular momentum changes by order unity in an orbital period, we draw periapse distances from the same distribution as is used for ``pinhole'' scattering \citep{Lightman:1977a}, e.g. $P(r_{\rm p,12}) \propto$~\mbox{constant}. $r_{\rm p,12,crit}/r_{\rm H,23}$ (Equation (\ref{eq:rpcrit})) ranges between 4.8 to 6.4 for $1 \leq q_{12} \leq 10^{3}$ for a circular orbit, but increases rapidly as $e_{12} \rightarrow 1$, growing to $10^{4}$ for $e_{12}$ values at the time of the final phases of the plunge. However, Equation (\ref{eq:rpcrit}) only specifies that a component of the triple system will be lost {\it eventually}, which could potentially be after many thousands of orbits. As an example, the Sun-Earth-Moon system, for which all three components are approximately circular and coplanar ($e_{12} \simeq 0$, $i_{23} \simeq 0$), is slightly unstable according to this criteria ($r_{\rm p,12}/r_{\rm p,12,crit} \simeq 1.16$), but has persisted for billions of orbits. Therefore, we expect that systems for which $r_{\rm p,12} \simeq r_{\rm p,12,crit}$ to almost never be lost in a single orbit, and systems for which $r_{\rm p,12} \simeq r_{\rm H,23}$ to usually become unbound with one passage.

As with the tertiary's orbit about the secondary, the secondary's orbit can also evolve by the emission of gravitational waves. Once this occurs, the secondary's orbit will circularize, and the production of SHS during the eccentric phase of the black hole merger will cease, with the emitted stars having velocities no greater than the orbital velocity of the secondary black hole (see Section \ref{sec:similarity}). Assuming that $e_{12} \sim 1$, the minimum periapse distance $r_{\rm p,\min}$ at which $T_{e}/\tau_{\rm GW,12} = 1$ (for $\alpha = 7/4$) is
\begin{equation}
\frac{r_{\rm p,\min}}{r_{\rm g,1}} \simeq 15 \left[\frac{(1+q)^2}{q^{8/5}}\right]^{1/7} \left(\frac{M_1}{10^8 M_\odot}\right)^{-1/7},
\end{equation}
corresponding to SHS that orbit the primary with $v/c~=~0.26 M_{1,8}$. Such stars are likely ejected at these velocities even from circular orbits \citep{Sesana:2006a}, comparable to the fastest stars produced during the eccentric merger phase which we detail in the next section. Additionally, if the secondary's orbit is truly plunging, its periapse distance may change by order unity over a single orbital period, this would yield at least a few orbits of moderate eccentricity for $r_{\rm p} < r_{\rm p,\min}$. In conjunction, these two effects will likely result in the production of SHS at a slightly reduced rate relative to what we present here (as we have ignored the effects of GW emission from the primary-secondary system), but for only the stars of the very highest velocities.

\section{Results}\label{sec:results}
In this section we present the results of our scattering experiments, and focus solely on the kinematic properties of SHS that result, neglecting observational prospects, which we cover in Section~\ref{sec:observations}.

\subsection{Fates of Tertiaries}
Our setup involve the passage of a two-body system (the tertiary in orbit about the secondary) by a more-massive primary on a near-parabolic orbit. Even for the most massive stellar tertiaries, the ratio of the tertiary mass to the secondary mass is minuscule, and therefore its presence does not significantly alter the trajectory of the primary or secondary MBHs. And because the orbit of the secondary is assumed to be plunging, the semimajor axis of the secondary is always significantly larger than its periapse distance, resulting in trajectories that are nearly identical in shape. Thus, if the semimajor axis distribution of stars about the secondary were a power-law, and no mechanism for stellar destruction existed, our results would not depend on $\tilde{a}$. It is the inclusion of the break in the power-law likely arising from stellar collisions, the tertiary's tidal disruption radius, and the secondary's Schwarzschild radius that eliminate this self-similarity, which effects the shares of outcomes at different $\tilde{a}$.

The production of SHS typically occurs when the orbit of a tertiary is dramatically altered by the tidal potential of the primary. The shrinkage of the secondary's Hill radius is not instantaneous, with the timescale being comparable to the secondary-tertiary orbital period $P_{23}$, and this means that the tertiary usually executes at least one orbit about the secondary during the encounter (this is evident in many of the examples shown in Figure~\ref{fig:examples}). During this time, the periapse distance of the tertiary about the secondary can change by factors of order unity, which can result in higher orbital velocities at the time of ejection, or can result in the tertiary's destruction when it crosses its tidal radius or the distance of the secondary's last stable bound orbit.

The top panel of Figure~\ref{fig:fate} shows the outcomes of our scattering experiments as a function of $\tilde{a}_{\min}$. For $\tilde{a} \sim 1$, these stars are already within a few $r_{\rm IBCO,2}$ of the secondary, and thus only a moderate perturbation by the primary is required to potentially knock them into the secondary. As the initial orbits about the secondary are thermal, changes in orbital angular moment of order unity are required to unbind them from the secondary, and this means that those encounters that would be likely to produce an SHS are also likely to result in the tertiary's destruction. As a result, small $\tilde{a}$ is characterized by encounters in which the tertiary is usually destroyed, and SHS production is relatively rare ($< 1\%$ of stars within this semimajor axis bin). This also reduces the number of objects that become bound to the primary after the encounter, as this outcome also requires changes in the tertiary's orbit of order unity.

Because the difference in mass between the primary and secondary is usually a factor of 10, the primary's IBCO $r_{\rm IBCO,1}$ and the periapse distance of the secondary $r_{\rm p,2}$ are comparable in size, occasionally resulting in the primary destroying the tertiary when it wanders into its IBCO. Because the typical mass of the primaries are $10^{9} M_{\odot}$, tidal disruption by the primary is extremely rare, as $r_{\rm IBCO,1}$ is usually several times larger than the tidal radius of MS stars. The secondary, being about an order of magnitude less massive, is at a mass where the IBCO and tidal radius are comparable, and thus tidal disruptions are more likely to be the means of destruction.

\begin{figure}
\centering\includegraphics[width=\linewidth,clip=true,angle=90]{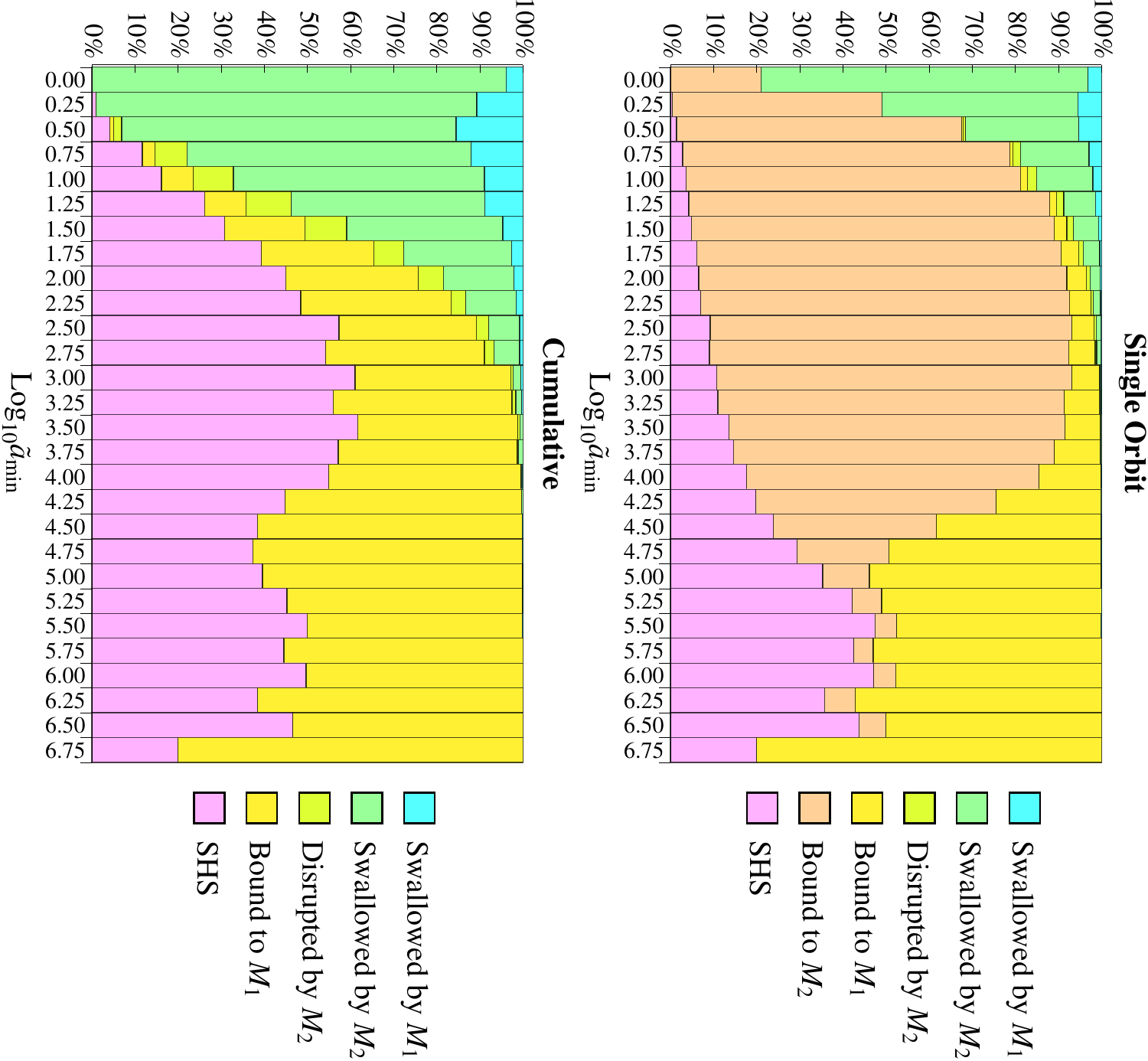}
\caption{Outcomes from individual scattering experiments as a function of semimajor axis bin, denoted by the minimum value of each bin $\log_{10} a_{\min}$. The top bar chart shows the direct outcome from each scattering experiment, and includes a ``bound binaries'' category in which the tertiary remains bound to the secondary after the encounter. The bottom bar chart assumes that any scattering event that did not result in either the removal or destruction of the tertiary will continue to be subject to repeated close periapse passages, eventually removing the star from its orbit about the secondary, with the probability of each outcome being the same as the single orbit case.}
\label{fig:fate}
\end{figure}

As $\tilde{a}$ increases to $\sim 10$, the number of objects destroyed by the secondary decreases significantly, as now orbits can be perturbed without much threat of destruction. The outcome of most encounters in this regime is that the tertiary remains bound to the secondary, although often with an alteration of its orbital parameters. SHS production in this range begins to become significant, with a few percent of stars becoming SHS per orbit. Because the orbit of the secondary is so eccentric when the most tighlty bound stars are perturbed, the specific orbital energy of the secondary about the primary is small, and as a result the outgoing velocity is mostly determined by the kick received via the Hills' mechanism.

For $\tilde{a}$ greater than a few $10^{4}$ the fraction of stars that become bound to the primary increases to become comparable to the fraction of SHS produced. This can be understood by comparing the specific binding energies of the primary-secondary system to the secondary-tertiary system. In the plunging scenario, the semimajor axis $a_{12}$ is nearly constant for all $\tilde{a}$ for fixed primary and secondary masses, so long as $r_{\rm p,2} \ll a_{12}$. Using Equation (\ref{eq:sigma}) we rewrite Equation (\ref{eq:astall}) in terms of $M_{1}$ and $q_{12}$, and find that the ratio of the specific binding energy of the primary-secondary system $\epsilon_{12}$ to the secondary-tertiary system $\epsilon_{23}$ is
\begin{equation}
\frac{\epsilon_{12}}{\epsilon_{23}} = 2.5 \left(\frac{\tilde{a}}{10^{5}}\right) \left(\frac{M_{1}}{10^{9} M_{\odot}}\right)^{1/2} \left(\frac{q_{12}}{10}\right)^{-4/5},
\end{equation}
showing that indeed the two energies become comparable for $\tilde{a} \sim 10^{5}$.

In the bottom panel of Figure~\ref{fig:fate} we show the fraction of final outcomes under the presumption that the per-orbit probabilities are unaffected by the stars' orbital histories. Because all other outcomes are mutually exclusive (a star swallowed by one of the MBHs cannot also be ejected), the cumulative encounter outcome is simply equal to the relative contributions of each outcome in the single case with the ``bound to $M_{2}$'' case being removed. \citet{Antonini:2010a} performed simulations of repeated encounters of binaries with a MBH and found that the rate of ejection did not increase significantly when considering multiple orbits where the periapse is held constant. However, in the case of a plunging secondary, the periapse distance shrinks on a timescale comparable to the orbital period, and thus the periapse does not remain constant from orbit to orbit. This makes stars progressively easier to remove on subsequent encounters as the secondary plunges deeper within the primary's potential.

\subsection{Properties of SHS}
As described in the previous section, the fate of a star in orbit about a plunging secondary is dependent on its distance from the secondary $\tilde{a}$ and its vulnerability to destruction, which depends mostly on the IBCO of the secondary and the tidal radius of the tertiary. If the star is ultimately ejected from the system, the distribution of outgoing velocities $v_{\infty}$ is also likely to depend on $\tilde{a}$ for two simple reasons: They are orbiting with larger velocities about the secondary to begin with, and some stars that would be ejected at a particular velocity may instead be destroyed by passing too close to the secondary.

\begin{figure}
\centering\includegraphics[width=0.9\linewidth,clip=true]{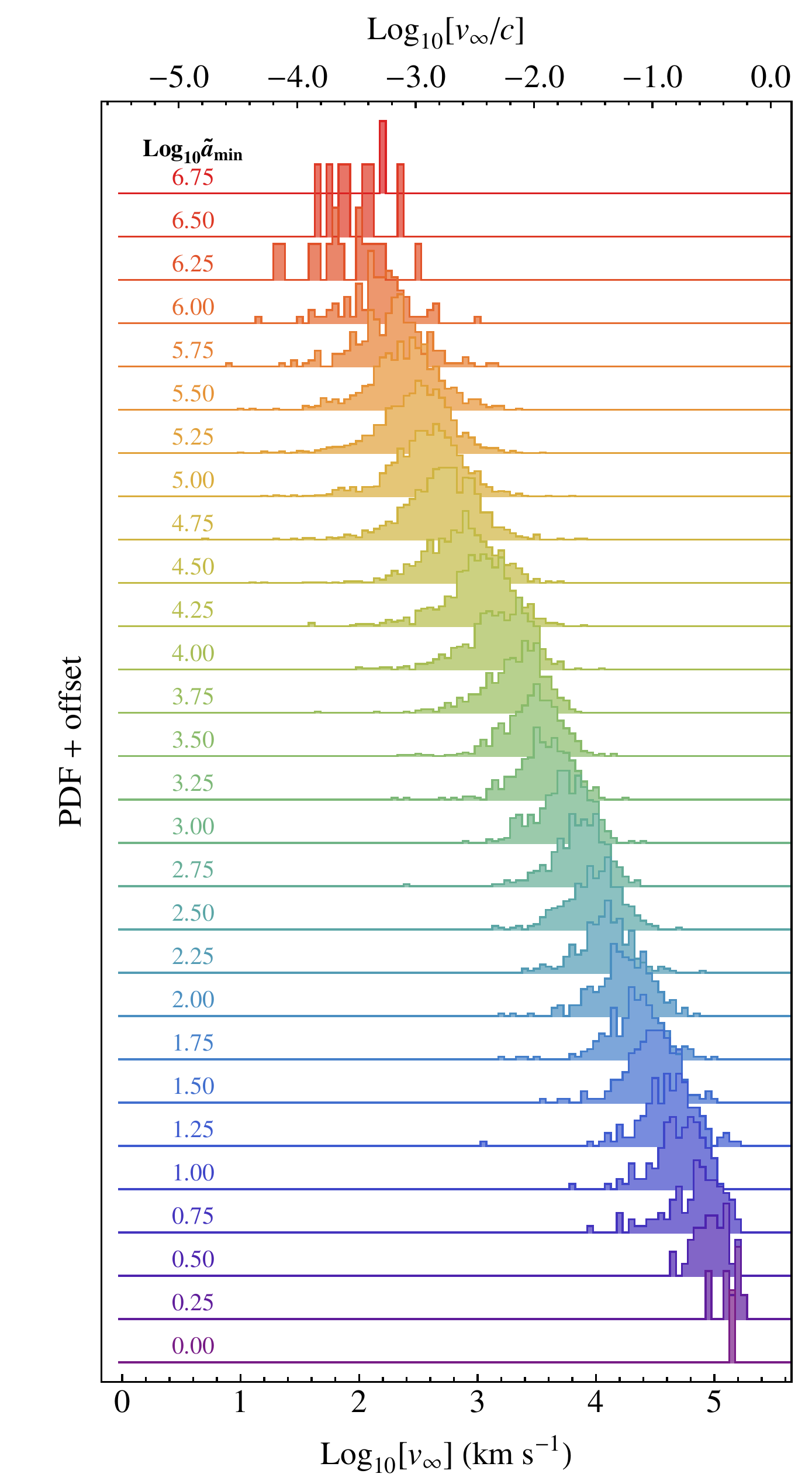}
\caption{Probability distribution functions of the asymptotic velocity $v_{\infty}$ of stars ejected from merging BMBHs. Each histogram shows the outcome of a set of scattering experiments for a restricted range of initial semimajor axes of the tertiary about the secondary, $a_{\min} < a_{23} < a_{\max}$, where $a_{\min}$ and $a_{\max}$ are spaced logarithmically in intervals of 0.25, with purple corresponding to $a_{\min} = r_{\rm IBCO,2}$ and red corresponding to $a_{\max} = 10^{6} r_{\rm IBCO,2}$. Each histogram shows the results of an independent scattering experiment composed of 4096 systems, where the parameter combinations are drawn as described in Section~\ref{sec:experiments}, and are plotted normalized to the bin with the most systems within each histogram.}
\label{fig:vinfpdf}
\end{figure}

\begin{figure}
\centering\includegraphics[width=0.9\linewidth,clip=true]{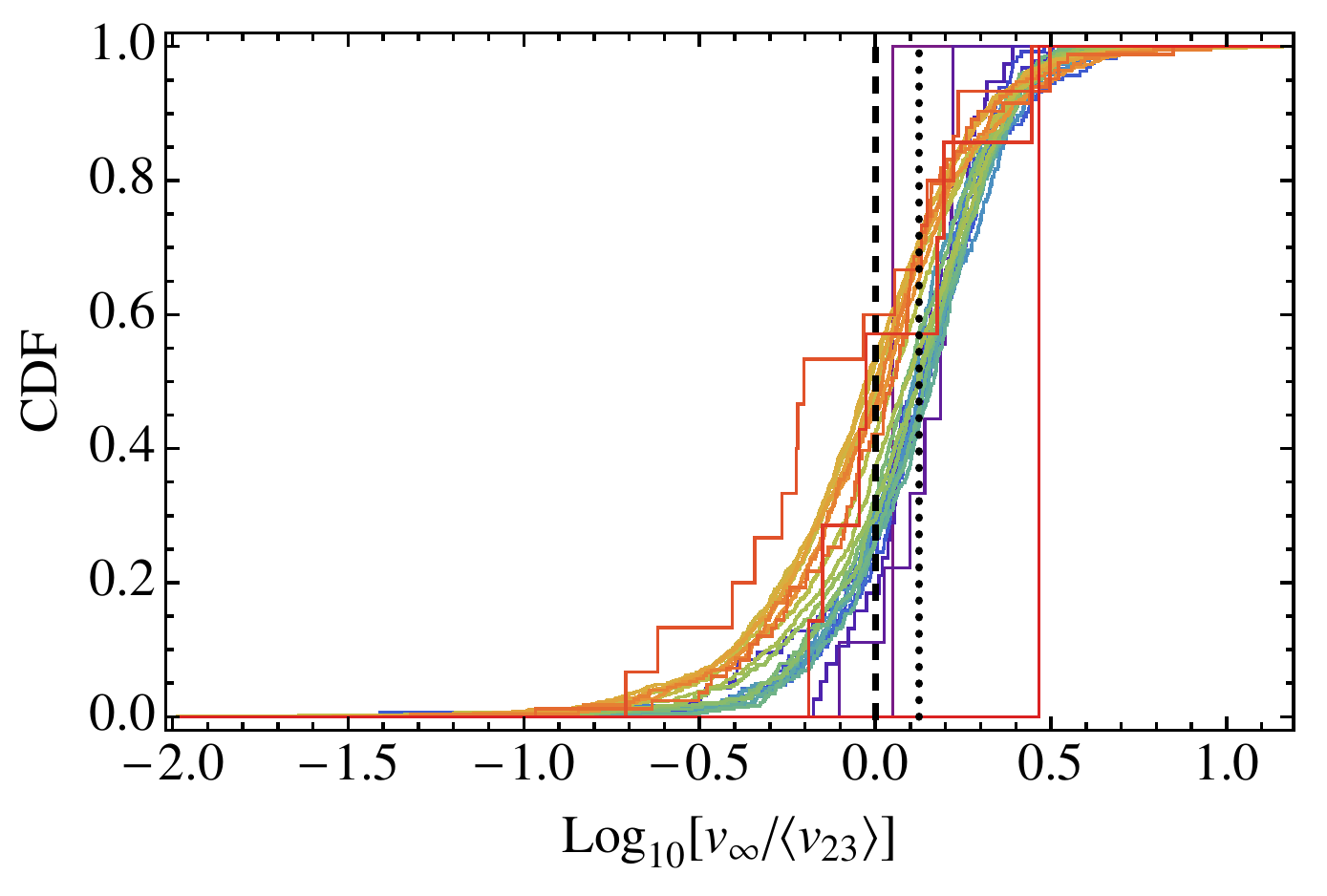}
\centering\includegraphics[width=0.9\linewidth,clip=true]{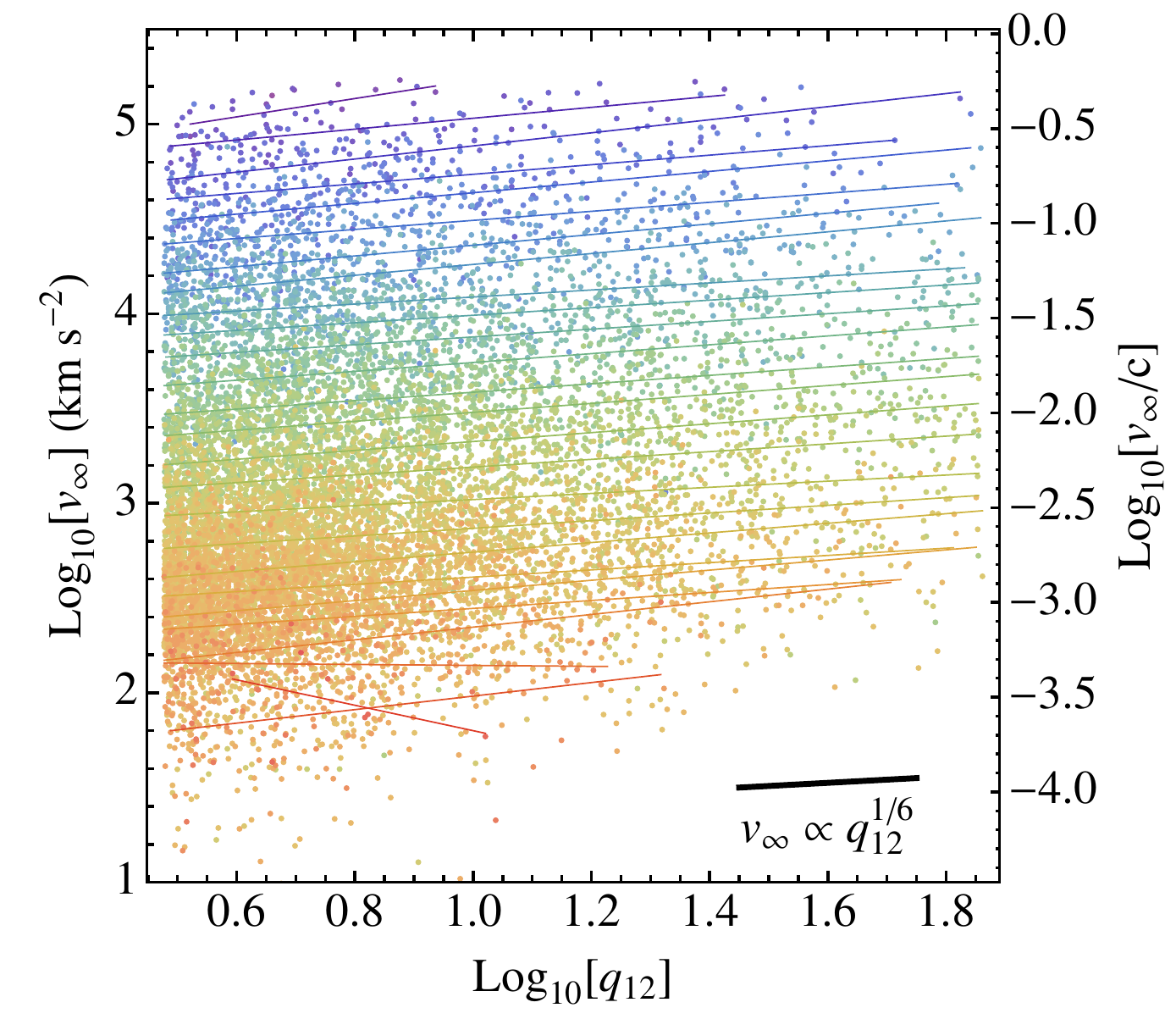}
\caption{The top panel shows cumulative distribution functions of the ratio of $v_{\infty}$ to the average velocity of the star's orbit about the secondary prior to its ejection, $\left<v_{23}\right>$, demonstrating the close relationship between the tertiary's original binding energy to the secondary and the mean ejection velocity. The black vertical dashed line shows what the mean value would be for $q_{12} = 1$, whereas the black vertical dotted line shows the enhancement to the mean value owing to the average mass ratio $q_{12} \simeq 5$ (Figure~\ref{fig:qplunge}). In the bottom panel, a scatter plot of $v_{\infty}$ is shown as a function of $q_{12}$, with each line showing the least-squares linear fit within a given semimajor axis bin, color-coded to the particular scattering experiment labeled in Figure~\ref{fig:vinfpdf}. The thick black line segment shows the expected $v_{\infty} \propto q_{12}^{\smash{1/6}}$ relationship; consistent with our linear fits, but with tremendous scatter. For extreme values of $\tilde{a}$ the slope is less well-defined (as an example $\log_{10} \tilde{a}_{\min} = 6.75$ includes only two systems), as our scattering experiments only produced a few SHS within these velocity bins.}
\label{fig:vratio}
\end{figure}

\begin{figure}
\centering\includegraphics[width=\linewidth,clip=true]{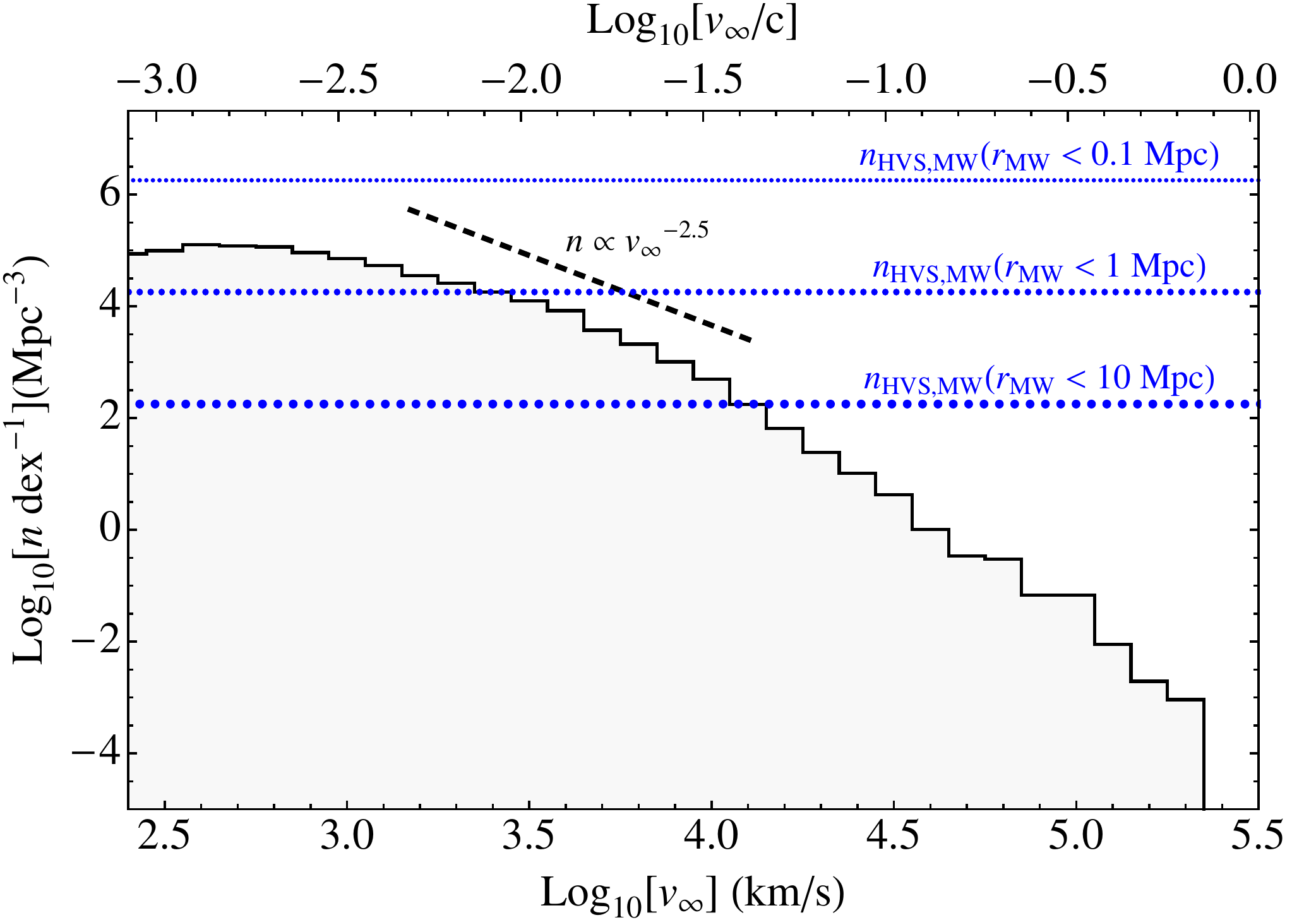}
\caption{Volume density $n$ as a function of $v_{\infty}$, ignoring the additional velocity required to escape the host galaxies' potentials. The dashed black line shows the slope found for the MBH mechanism \citep{Sesana:2007b,Sherwin:2008a}. The dotted blue lines show the average density of HVS produced by the MW within a certain distance of the MW $r_{\rm MW}$, where the distances are as labeled in the figure, where we have assumed that HVS have velocities equal to 3,000 km s$^{\smash{-1}}$.}
\label{fig:histogram}
\end{figure}

\subsubsection{Velocity Distribution}
All stars that become unbound to the secondary inevitably acquire a positive excess velocity relative to it, but this excess velocity is not necessarily sufficient for the star to escape the primary as the secondary is bound to it. This leads to the production of a population of stars that remain bound to $M_{1}$ ($\epsilon_{13} < 0$), and a population unbound from $M_{1}$ ($\epsilon_{13} > 0$), with values spanning the two extremes. What determines the amount of spread are the initial conditions of the secondary-tertiary system, primarily its phase and orientation. Because the values span both positive and negative values, this means that SHS with positive $\epsilon_{13}$ can be ejected with velocities that are arbitrarily close to zero, even if the time-averaged orbital velocity of the tertiary about the secondary $\left<v_{23}\right>$ (prior to removal) is significant.

In Figure~\ref{fig:vinfpdf} we show the output distributions of $v_{\infty}$ as a function of $\tilde{a}$, with each histogram summing across the experiments drawn from the expected triple configurations, as described in Section~\ref{sec:experiments}. As each histogram is produced via an identical number of experiments, the small number of systems contributing to the distributions at both large and small $\tilde{a}$ demonstrates that producing SHS for these extreme semimajor axis values is difficult. The output distributions are also visibly Gaussian, and centered about a particular $v_{\infty}$ for each $\tilde{a}$ (Figure~\ref{fig:vratio}, top panel), although this value is somewhat in excess of $\left<v_{23}\right>$. This excess is mostly due to the fact that the secondary is in motion at the time of the tertiary's escape, which gives an additional kick to the tertiary on top of its time-averaged velocity, scaling as $q_{1,23}^{\smash{1/6}}$ \citep{Hills:1988a}. Even though the orientation and phase of the binary play a large role in determining $v_{\infty}$ for a particular system, the fit to datapoints clearly shows this relationship (Figure~\ref{fig:vratio}, bottom panel).

The combined probability distribution determined by properly normalizing the distributions presented in Figure~\ref{fig:vinfpdf} is shown in Figure~\ref{fig:histogram}. We find that the slope of the total number of objects $n$ is similar to that found for stars that are originally bound to the primary and are scattered by the incoming secondary \citep[see Figure~1 of][]{Sesana:2007b} for objects with $v_{\infty} \sim$ several thousand km s$^{-1}$, but that the slope steepens for higher velocities. This is likely because we assume that the radial profile of stars flattens out interior to the distance at which stars are frequently destroyed by stellar collisions. When compared to the local density of HVS, which we calculate assuming an average velocity of 3,000 km s$^{-1}$ \citep{Bromley:2006a} and a production rate of $10^{-4}$ yr$^{-1}$ \citep{Yu:2003a}, a nearly equal number of SHS occupy a volume within 1 Mpc of the MW as HVS. This distribution does not take into account the fact that SHS must climb out of the potential wells of their host galaxies, and as we will describe in Section~\ref{sec:observations}, this effect can significantly reduce the local population of low-velocity SHS.

\subsubsection{Dependence on Initial Orientation and Phase}
In the left panel of Figure~\ref{fig:angle0} we show the CDF for the $z$-component of the secondary-tertiary system's angular momentum vector \smash{$\vec{J}$}, where $z$ is defined to be perpendicular to the primary-secondary's orbital plane. Even though we tend to draw smaller $r_{\rm p,12}$ (Equation (\ref{eq:rpcrit})) for retrograde systems (i.e. $J_{z} < 0$), it is clear that prograde systems are more likely to result in ejection, with approximately 60\% of all SHS originating from systems in which $J_{z} > 0$. The trend is most pronounced for the fastest SHS, with 75\% of SHS being produced from prograde systems when $\tilde{a} < 10$. As $\tilde{a}$ increases to $\sim 10^{4}$, the number of systems ejected depends less on whether the system is prograde or retrograde; we speculate that this is because the primary-secondary orbit becomes more circular with increasing $\tilde{a}$, and thus each secondary-tertiary pair orbits about one another a few times (as opposed to roughly once in a parabolic encounter) while in the vicinity of the primary's tidal field, averaging out the tertiary's orbital motion.

\begin{figure}
\centering\includegraphics[width=0.9\linewidth,clip=true]{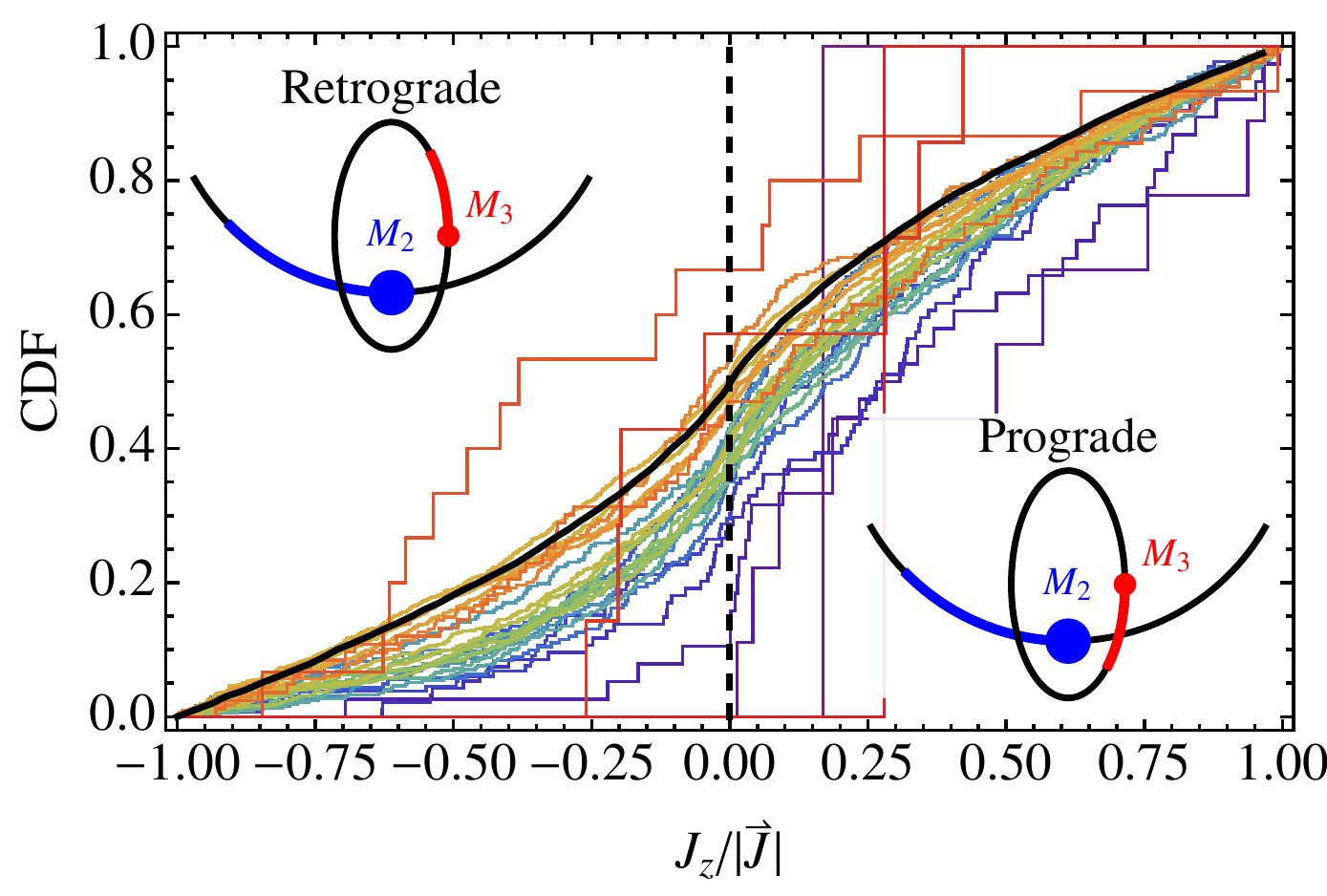}\\
\includegraphics[width=0.9\linewidth,clip=true]{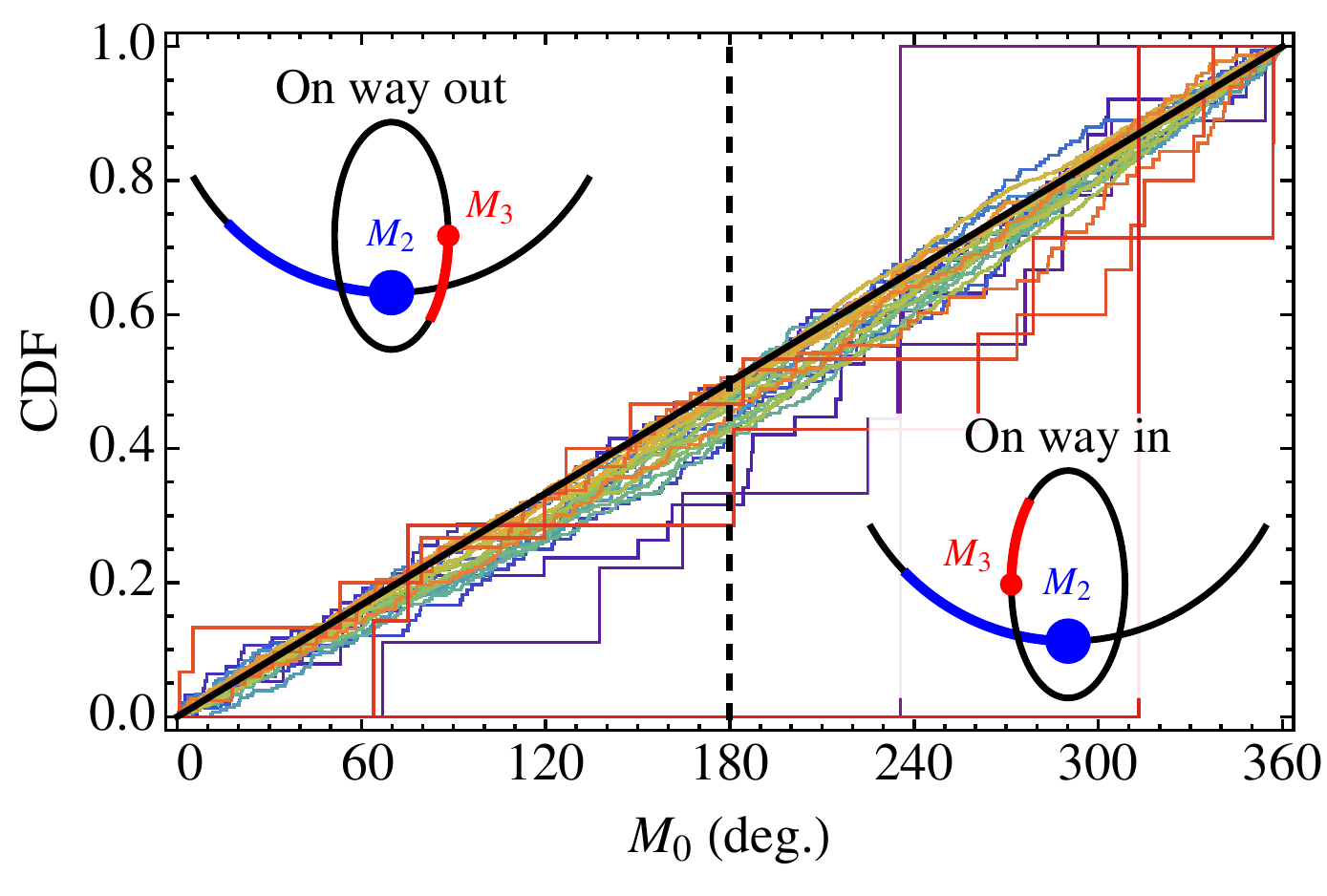}
\caption{Cumulative distribution functions of the $z$-component of the initial angular momentum vector $\vec{J}$ of the secondary-tertiary orbit relative to the primary-secondary plane, and initial mean anomaly $M_{0}$ of stars that wound up being ejected from their BMBH system. Each CDF is color-coded to the particular scattering experiment labeled in Figure~\ref{fig:vinfpdf}. The solid black curves shows the input distributions assumed for each variable, with $J_{z}$ being randomly drawn from a 2-sphere and $M_{0}$ being drawn from a uniform distribution. The dashed vertical line shows a separation of retrograde versus prograde orbits in the left panel, and orbits in which the tertiary is moving toward apoapse versus periapse in the right panel.}
\label{fig:angle0}
\end{figure}

In addition to a dependence on the angles of the two systems to one another, the kick received can also depend on the tertiary's position within its orbit near the time of periapse. In our simulations, each run was performed an exact integer multiple of the secondary-tertiary system's orbit period prior to periapse, with the initial mean anomaly $M_{0}$ being drawn uniformly, where $M_{0} = 0^{\circ}$ (or $360^{\circ}$) corresponds to the tertiary being at periapse. As a result, $M_{0}$, in conjunction with the orientation of the orbit, directly relates to the position of the tertiary when the secondary reaches periapse, although not perfectly as the orbit is subject to some perturbation by the primary over the course of the orbit. The right panel of Figure~\ref{fig:angle0} shows that ejections are disfavored when the tertiary has recently crossed periapse and is on its way toward apoapse ($0^{\circ} < M_{0} < 180^{\circ}$), and are favored when the tertiary is returning to periapse ($180^{\circ} < M_{0} < 360^{\circ}$). This is as expected; when the tertiary approaches periapse it is in the process of accelerating, so the additional acceleration provided by the primary in this phase can unbind the tertiary to a greater degree than the case where the tertiary is deccelerating. 

\subsubsection{Dependence on the Masses of the Three Bodies}

\begin{figure}
\centering\includegraphics[width=0.9\linewidth,clip=true]{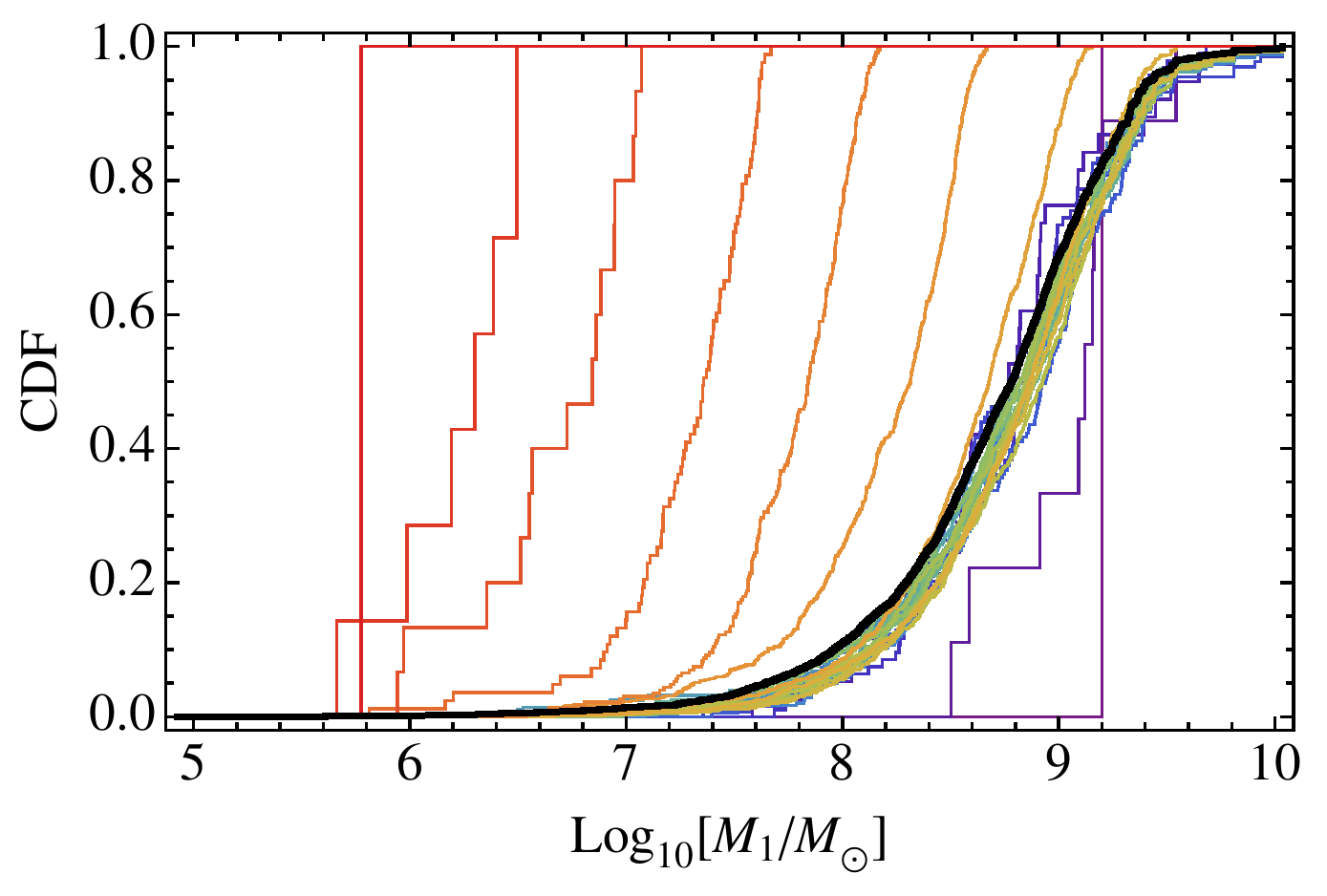}\\
\includegraphics[width=0.9\linewidth,clip=true]{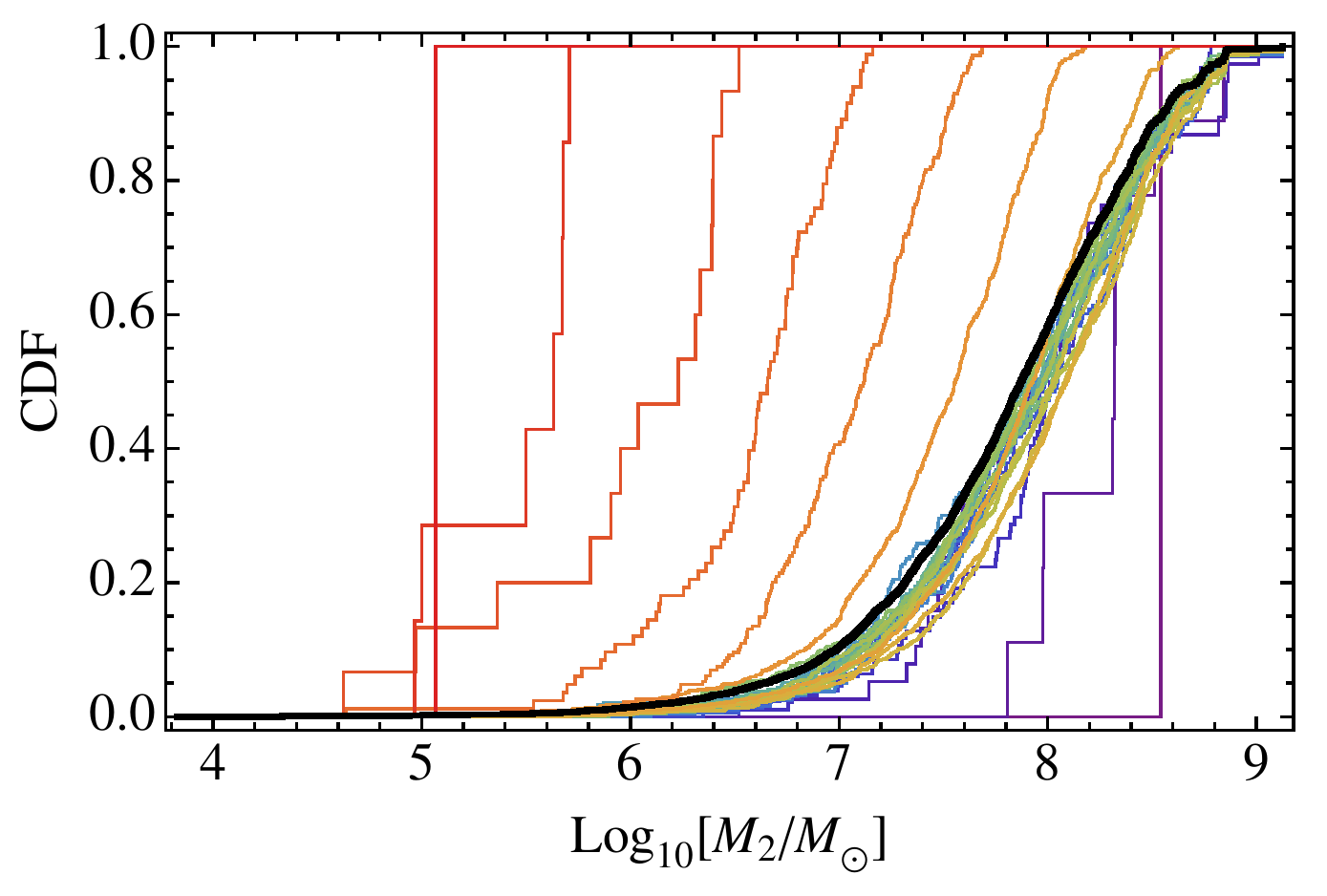}\\
\centering\includegraphics[width=0.9\linewidth,clip=true]{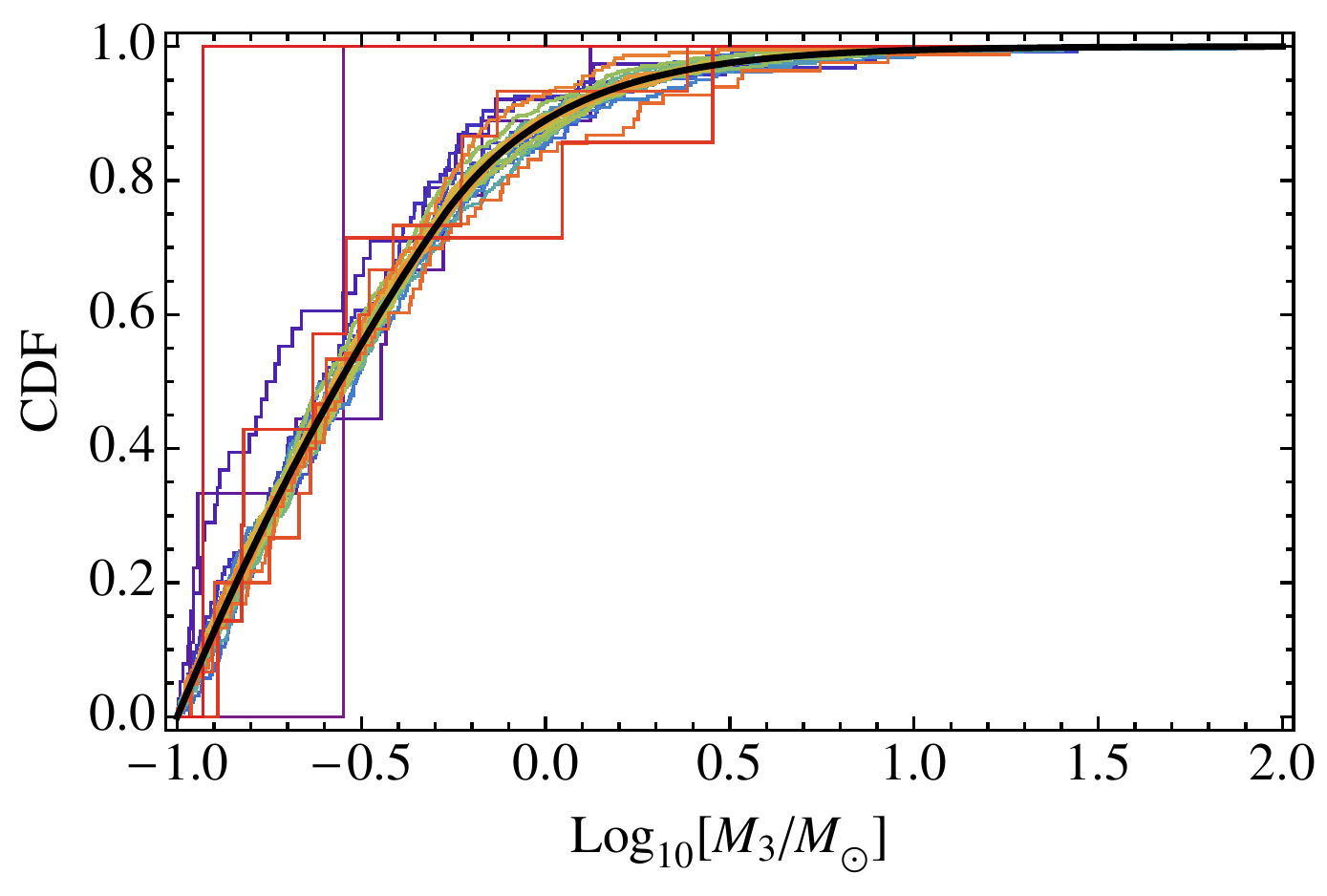}
\caption{Cumulative distribution functions of the primary, secondary, and tertiary masses, and the ratio of the secondary to the primary mass $1/q_{12}$. Each CDF is color-coded to the particular scattering experiment labeled in Figure~\ref{fig:vinfpdf}. The thick black curve in each panel shows the initially assumed distributions for each mass, as described in Section~\ref{sec:experiments}.}
\label{fig:mcdf}
\end{figure}

Figure~\ref{fig:mcdf} shows CDFs for the masses of the primary, secondary, and tertiary bodies. Immediately apparent in the primary and secondary distributions (top two panels) is that most of the large $\tilde{a}$ systems originate from low-mass black holes, this simply reflects the fact that the sphere of influence is smaller than $\tilde{a} r_{\rm IBCO, 2}$ for large black holes. For $\tilde{a}$ in which all black hole masses are included, a slight preference toward higher black hole mass is apparent for both the primary and the secondary as compared to the input distributions, with the exception of large $\tilde{a}$ which can only originate from small black holes for which the sphere of influence is much larger than the IBCO.

For the tertiary, whose mass is tremendously smaller than the primary and secondary, is for all intents and purposes a test particle; no statistically significant dependence on its mass is apparent (Figure~\ref{fig:mcdf}, lower panel). One might expect that the inclusion of the tidal disruption radius would preferentially destroy stars with a lower density, but the vast majority of systems that produce SHS have masses in excess of $10^{8} M_{\odot}$, a mass for which the IBCO lies exterior to the tidal radius for most MS stars, as is apparent from the small fraction of tidal disruptions accounted for in Figure~\ref{fig:fate}.

\subsubsection{Dependence on Eccentricity of the Tertiary}

\begin{figure}
\centering\includegraphics[width=0.9\linewidth,clip=true]{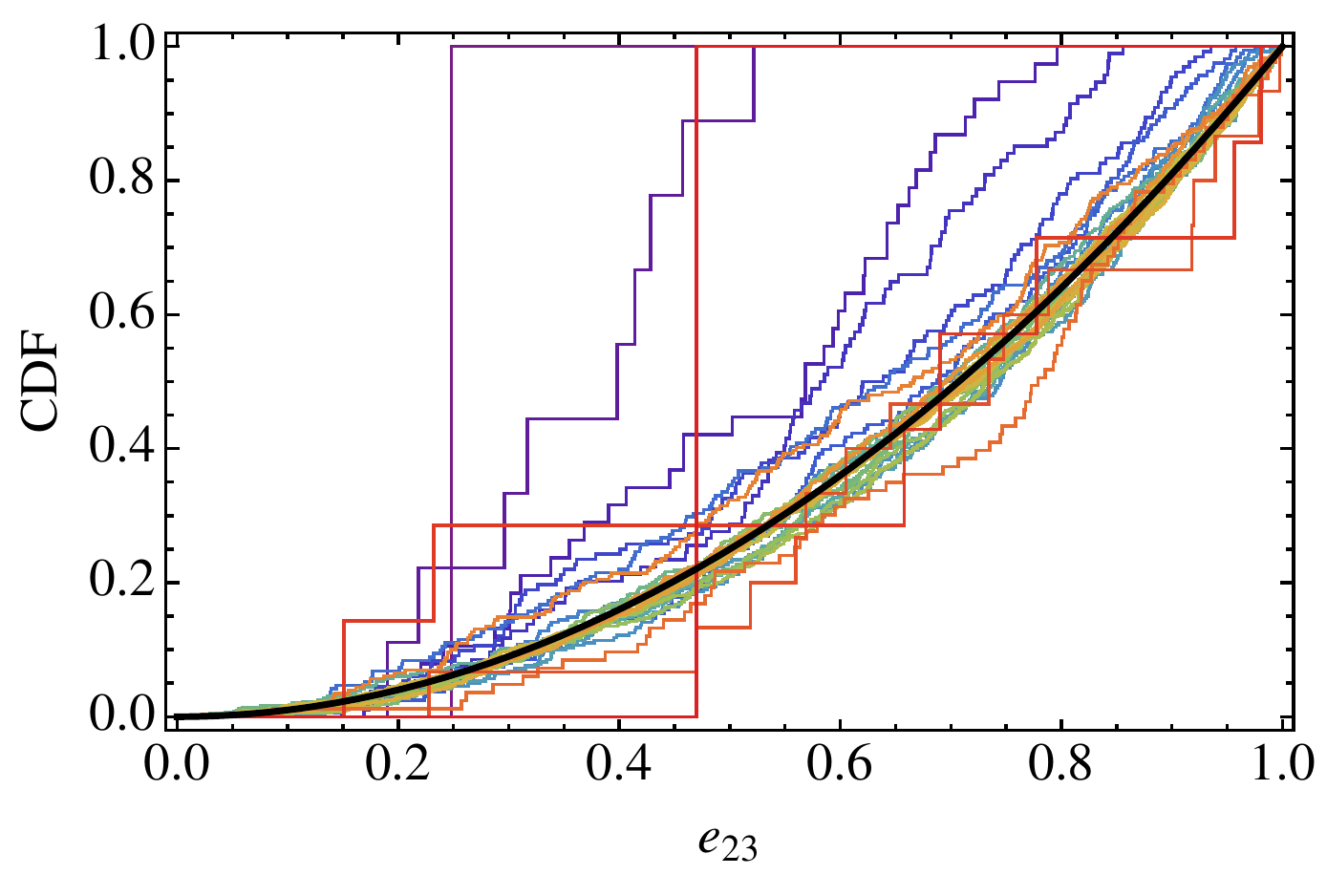}
\caption{Cumulative distribution function of the initial eccentricity of the star about the secondary $e_{23}$. Each CDF is color-coded to the particular scattering experiment labeled in Figure~\ref{fig:vinfpdf}, and the thick black curve shows the initially assumed thermal distribution, $P(e) \propto e$.}
\label{fig:ecdf}
\end{figure}

The stars that orbit our own galactic center exhibit a wide range of eccentricities, but the inner cluster of stars seems to be well-described by a thermal distribution \citep{Gillessen:2009b}. While some stars were likely deposited via the disruption of stellar binaries resulting in much higher initial eccentricities \citep{Madigan:2011a}, we assumed that the stars in orbit about the secondary are drawn from a thermal distribution because of the strong relaxation effect of the repeatedly passage of the primary black hole \citep{Perets:2007a}, which is likely to destroy any pre-existing coherent structures in angular momentum space.

Figure~\ref{fig:ecdf} shows the cumulative distribution function of SHS as a function of the secondary-tertiary eccentricity $e_{23}$. While our experiments for large $\tilde{a}$ show no statistically significant deviation from the input distribution, small $\tilde{a}$ does show that more-circular systems are preferably accelerated. The reason for this is that tertiaries in highly eccentric orbits are more likely to be destroyed after being perturbed by the primary, as their periapses are initially closer to the swallowing radius $r_{\rm IBCO,2}$. For systems that are not swallowed, we find no statistically significant trend between $e_{23}$ and $v_{\infty}$. Therefore, eccentricity distributions that are biased toward higher eccentricities than thermal are likely to result in a reduction of the number of SHS produced.

\subsubsection{Beaming of Outgoing SHS Trajectories}\label{sec:beaming}

\begin{figure*}
\centering
\begin{minipage}[b]{0.45\linewidth}
\centering
\includegraphics[width=0.8\linewidth,clip=true]{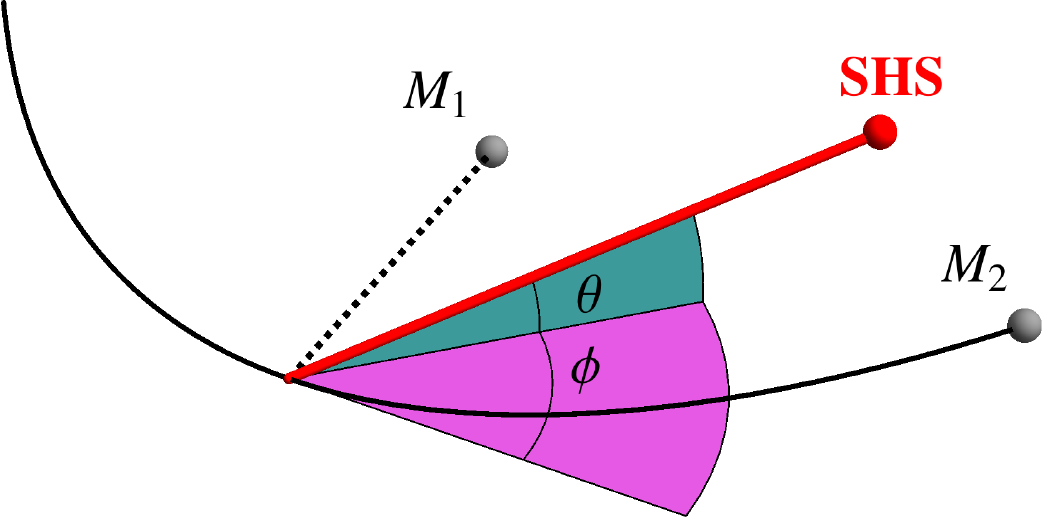}\\
\includegraphics[width=0.8\linewidth,clip=true]{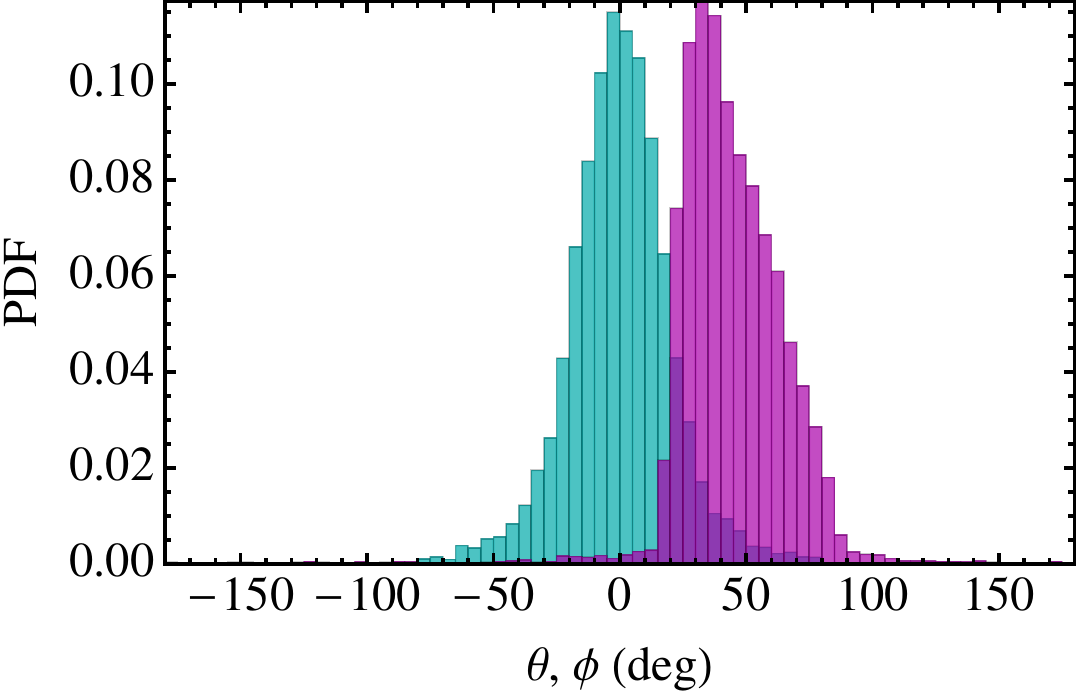}
\end{minipage}
\begin{minipage}[b]{0.45\linewidth}
\centering
\includegraphics[width=\linewidth,clip=true]{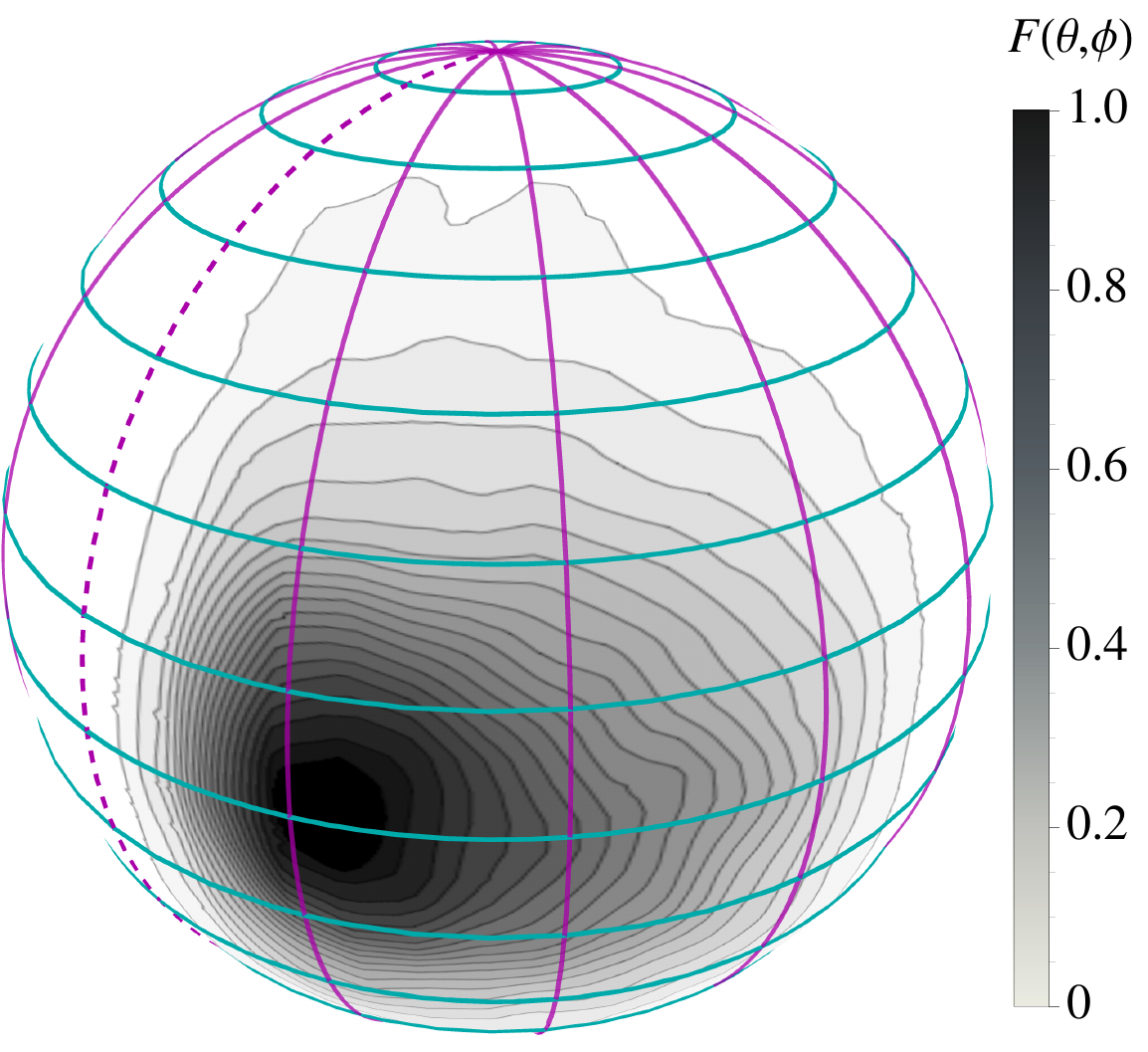}
\end{minipage}
\caption{Distribution of velocity vector angles for SHS. Cumulative fraction $F(\theta, \phi)$ of ejected stars with $v_{\infty} > 0$. The diagram in the upper left shows the positions of the three objects (primary black hole, labeled $M_{1}$, secondary black hole, labeled $M_{2}$, and tertiary star, labeled SHS) shortly after the ejection of an SHS, where $\phi$ is the polar angle and $\phi = 0$ is defined to be the angle of $M_{2}$'s velocity vector at periapse, and the azimuthal angle $\theta$ is measured from the orbital plane. The single-parameter histograms in the bottom-left panel and the 2D histogram mapped to the surface of a sphere show that SHS are preferentially "beamed" in a cone centered at $\theta = \pi/4$ and $\phi = 0$ \citep[see][]{Zier:2001a,Sesana:2007a}.}
\label{fig:ejangles}
\end{figure*}

In the center-of-mass frame of the triple system, the motion of the secondary and tertiary both follow a nearly-parabolic elliptical orbit about the primary. Because the tertiary is unbound from the secondary near periapse, its position at this point defines the apex of a cone of outgoing trajectories. The unbound objects leave the system on hyperbolic orbits, where their eccentricity is directly related to the excess orbital velocity and the secondary's periapse distance,
\begin{equation}
e = \sqrt{1 + \frac{2 v_{\infty}^{2} r_{\rm p,12}}{G M_{1}}}\label{eq:hypere}.
\end{equation}
As $r_{\rm p,12} = a_{23} q_{12}^{1/3}$, and $v_{\infty}^{2} \simeq v_{23}^{2} q_{12}^{1/3} \simeq G M_{2} q_{12}^{1/3} / a_{23}$, Equation (\ref{eq:hypere}) becomes
\begin{equation}
e \simeq \sqrt{1 + 2 q_{12}^{-1/3}}.
\end{equation}
This in turn is related to the angle of the outgoing velocity vector by $\phi = \pi/2 - \arccos (1/e)$, where $\phi$ is measured from a vector parallel to the secondary's motion at periapse (Figure~\ref{fig:ejangles}, upper left panel) and is equal to $\pi/2$ for a parabolic orbit, and plugging in the median mass ratio $q_{12} \simeq 5$ (Figure~\ref{fig:qplunge}), we find that the average ejection should have $\phi~\simeq~40^{\circ}$.

In the lower left panel of Figure~\ref{fig:ejangles} we show histograms of $\phi$ (longitude), as defined above, and $\theta$, the azimuthal angle (latitude). We find that indeed the average ejection $\phi$ has a value of around $40^{\circ}$, although with some considerable scatter. Given our isotropic angle distribution assumption, it is not surprising that our $\theta$ distribution is symmetrical about zero, and with a scatter comparable to the average $\phi$ value, with latitudes near the orbital plane being preferred \citep{Sesana:2006a}. The fraction of solid angle that the SHS are beamed into is rather small, with approximately half the SHS emanating from $\sim 5\%$ the area of the full sphere (Figure~\ref{fig:ejangles}, right panel).

If the secondary's orbital evolution is plunging, this implies that many stars are lost from the secondary nearly simultaneously, perhaps even on the same orbit. If this is the case, the beaming described above would cause stars from these merger events to be confined to a small cone, which would enhance the number of SHS originating from particular merger events in which the cone is pointed at us, at the expense of many merger events beaming their stars in directions away from the Earth. And because the majority of SHS are produced by the most-massive black hole mergers, this may mean that a select few merging MBHs would be responsible for the lion's share of the local SHS distribution.

In Figure~\ref{fig:m87} we show an illustrative example where a MBH merger occurred in M87 at a time 3 Gyr prior to the present and whose cone was beamed directly at the MW. For this figure, we assume that stars within a distance 2 Mpc of the MW are detectable, and additionally that this one merger event is responsible for all SHS in this velocity range, as might be expected if the beaming of SHS is ubiquitous. The figure shows that if SHS are indeed beamed, and the three-dimensional velocity vectors of multiple SHS can be determined, those stars with similar total velocities are likely to originate from the same galaxy, whose location can be determined by tracing their paths back to a common origin. This prospect is discussed in more detail in Section~\ref{sec:ident}.

\begin{figure}
\centering\includegraphics[width=\linewidth,clip=true]{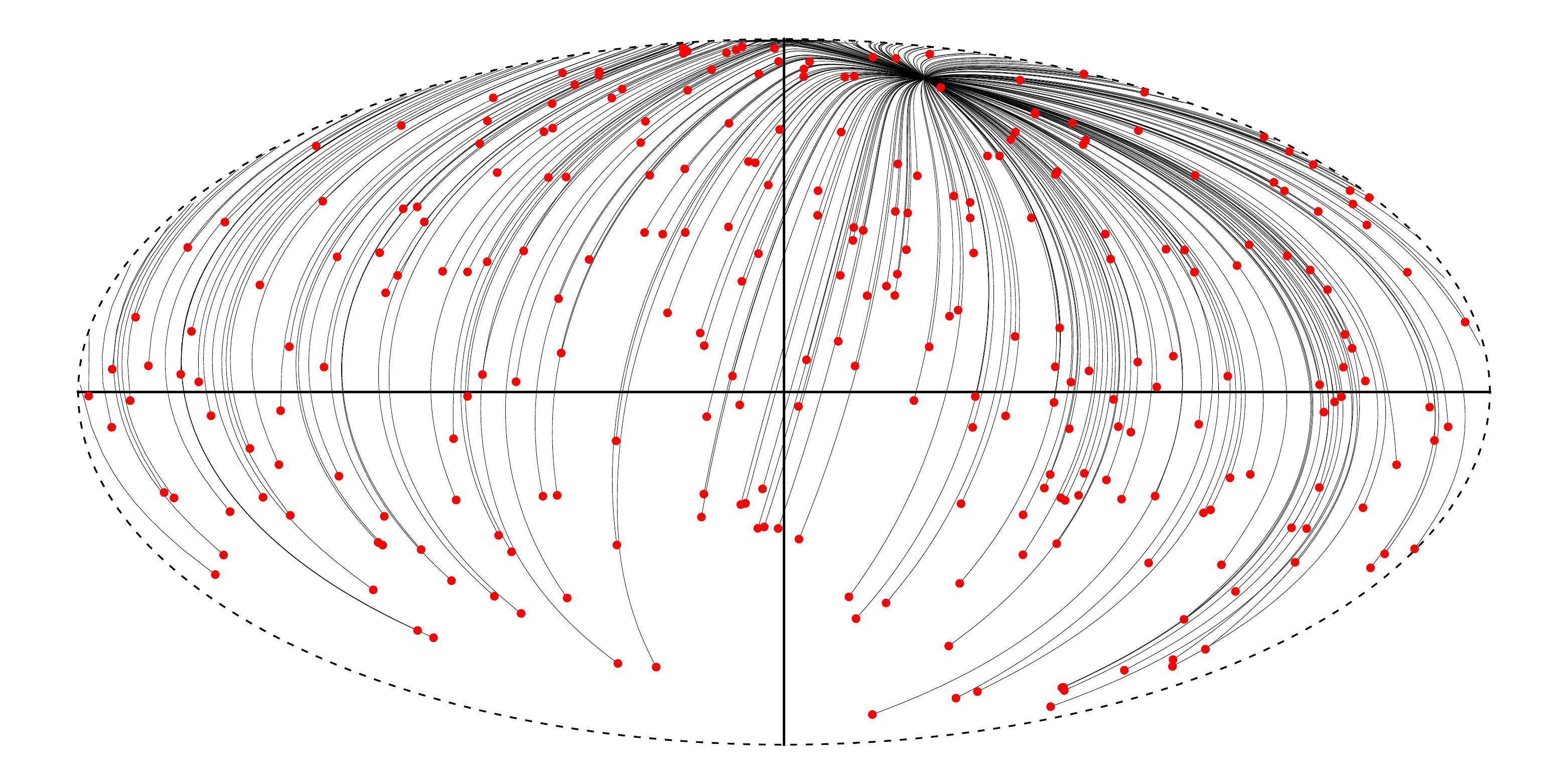}
\caption{Trajectories of SHS within 2 Mpc of the MW under the presumption that all originate from a single merger event 3 Gyr ago in M87. The plot shows the present locations of SHS in a Hammer projection of galactic coordinates, where the thin black lines trace their path back to their galaxy of origin. Deflection by either intervening substructure or the MW itself is not taken into account.}
\label{fig:m87}
\end{figure}

\section{Semi-relativistic hypervelocity binaries}\label{sec:shb}
A unique feature of the SHS ejection process is that the tertiary need not be a single star, but itself can be replaced by a binary. For tight binaries, the tidal disruption radius of the binary is only slightly larger than the tidal disruption radius of the star, and because binaries have a chance to survive even when passing deep within a tidal potential \citep{Sari:2010a}, they can sometimes survive even in extremely strong tidal encounters.

\begin{figure*}
\centering
\includegraphics[width=0.7\linewidth,clip=true]{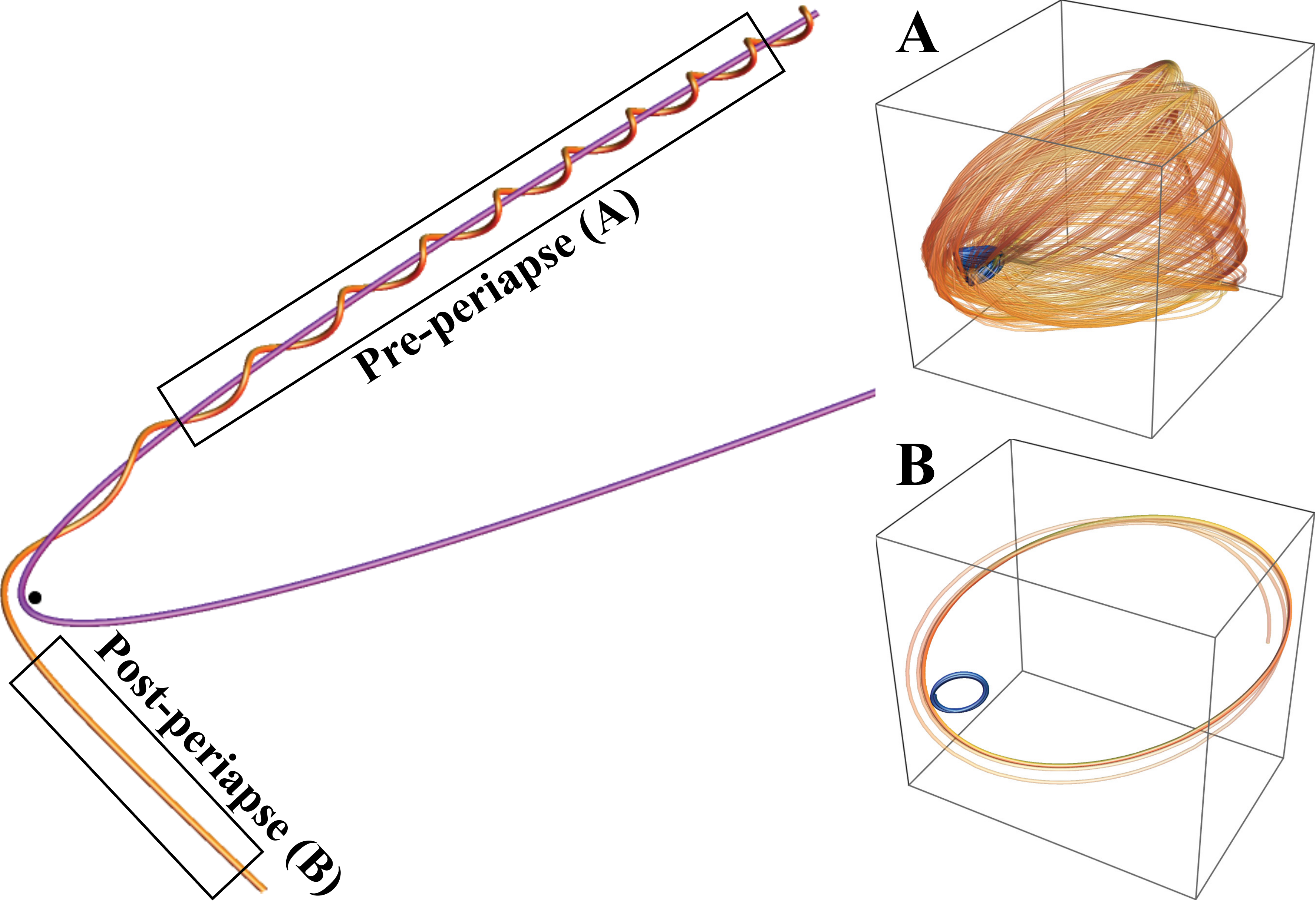}
\caption{Example of a BMBH encounter in which an SHB is produced. The left panel shows the orbital trajectory of the secondary (purple path) about the primary (black dot), and the tertiary-quartary's path about the secondary (orange path). Two regions are highlighted in the left panel and shown as insets in the tertiary-quartary frame, where the tertiary is colored orange and the quartary is colored blue: A pre-periapse phase (A) and a post-periapse phase (B). Before periapse, the tertiary-quartary system is on an eccentric orbit about the secondary and demonstrates a complicated eccentric Kozai excitation in which eccentricity and inclination are exchanged. The action of the secondary passing the primary results in the unbinding of the tertiary-quartary system from the secondary, at which point the eccentric Kozai ceases to be driven, and the stellar binary settles into a stable orbit.}
\label{fig:shb}
\end{figure*}

The survival of a multiple system undergoing the Hills mechanism has been studied in the context of triple stellar systems \citep{Perets:2009c} and for planetary systems orbiting one of the two stars \citep{Ginsburg:2012a}. Hypervelocity binaries have also been shown to be produced by a chance scattering via the inspiral of an intermediate mass black hole, similar to the mechanism we describe here, but where the binary is originally bound to the primary MBH \citep{Levin:2006a,Lu:2007a,Sesana:2009a}. In these papers, either the primary-secondary system is treated as being on fixed orbits \citep{Lu:2007a}, or the tertiary-quartary binary is not modeled explicitly \citep{Sesana:2009a}. Here we present the first set of simulations in which no approximations are made, and the orbits of all four bodies are solved for explicitly, however it should be noted that the assumption that the primary-secondary system remains unperturbed is a perfectly valid one if one is only concerned with the fate of the outgoing stellar binary. By evolving the orbits of all four bodies explicitly, we are able to calculate the energy lost by the primary and secondary in the ejection, as we do in the three-body case, which enables us to guarantee that the total energy and angular momentum of the system is conserved.

We show the production of an example semi-relativistic hypervelocity binary (SHB) in Figure~\ref{fig:shb}. The orbit of the stellar binary, which we refer to as the tertiary-quartary system, is selected such that its periapse is beyond the distance specified by Equation (\ref{eq:rpcrit}), where all variable indices are incremented by one (as we are considering two objects in orbit about the secondary). While satisfying this criterion guarantees that the system will not be disrupted, it doesn't prohibit secular evolution of the orbit of the secondary-tertiary-quartary triple system, and the upper right panel of Figure~\ref{fig:shb} shows that a clear eccentric Kozai excitation \citep{Kozai:1962a,Lithwick:2011a,Naoz:2013a,Li:2014a} is taking place prior to the secondary's periapse about the primary. Generically, the orbits of stars (and binaries) about an MBH have non-zero eccentricities, and so long as the two stars are themselves not interacting tidally, they too will have a range of eccentricities. As a result, eccentric Kozai excitations of varying magnitude are likely common pre-ejection states for SHB, especially for the fastest examples where the minimum separation is limited by the physical size of the stars in the binary.

Because more SHS are produced by the mergers of near-equal mass black holes, the separation between the tertiary-quartary and the secondary is often times comparable to the primary-secondary distance at periapse (see Figure~\ref{fig:examples}), meaning that the tertiary-quartary system can in some cases be disrupted by the primary if it lies close to it at periapse. In the example presented in Figure~\ref{fig:shb}, the tertiary-quartary system is further from the primary than the secondary, being on the ``outside track,'' and thus experiences tidal forces upon it from the primary that are weaker than those imposed by the closer secondary, but this is not guaranteed for all encounters. Just as in the single star case, the tertiary-quartary system can also be driven closer to the secondary, which can also disrupt it; again our example shows a case in which this does not occur. Figure~\ref{fig:fate} showed that very few single stars were tidally disrupted by either black hole (most of those that were destroyed were swallowed whole), this is likely to be different for SHB that have a tidal radius that is at least a factor of a few larger than that of single stars.

After the secondary reaches periapse, and if the tertiary-quartary system survives the tidal forces of both black holes, the system can be ejected at velocities that are comparable to SHS velocities. Of course since the stellar binary leaves the vicinity of both black holes, it settles into a simple elliptical orbit that is subject to no external perturbation. The final eccentricity and semi-major axis distributions of these objects is not calculated here, but is likely to be more extreme relative to binaries formed via other processes. As a result, such objects may merge more-frequently than the average stellar binary, and perhaps the population of SHB may be somewhat enhanced in merger products relative to the field.

\section{Observations}\label{sec:observations}

Once accelerated, SHS travel across the universe with little influence from self-gravitating structures, with the escape velocities from the most massive structures rarely exceeding a few $10^{3}$ km s$^{-1}$ \citep[as an example, the Phoenix cluster, one of the most massive known clusters, has an escape velocity of $\sim$~3,000~km~s$^{-1}$,][]{McDonald:2012a}. As star formation activity peaks at $z \sim 2$, and low-mass stars are more common than high-mass, the stars that survive their journey such that they are potentially observable are those with lifetimes in excess of $\sim 10$ Gyr, similar to the MS lifetime of the Sun. Stars with masses $\lesssim 1 M_{\odot}$ can remain on the MS for far longer than a Hubble time, but can be  significantly dimmer than the Sun at solar metallicity. Those stars that evolve into giants are significantly brighter and thus easier to detect, however the giant phases are typically short ($\lesssim$~1 Gyr). Because most SHS are produced when the star formation and galaxy merger rates are highest, and because the majority of detectable SHS are those with MS lifetimes comparable to the travel time, the typical age of detectable SHS will reflect the time at which they were ejected.

\subsection{The unbound SHS population}
To determine the population of SHS that would be detected by a survey, we use the results from our SHS scattering experiments as inputs into a second Monte Carlo calculation. We assume that the initial velocity distribution of SHS is universal and has no dependence on system properties, as suggested by our scattering experiments in which most of the dependence is on $\tilde{a}$ and $q_{12}$, both of which are scale-free. Random positions are drawn within a sphere of radius $r_{\rm Virgo} = 16~{\rm Mpc}$, where $r_{\rm Virgo}$ is approximately the distance to the Virgo cluster, the maximum distance to which future surveys will be able to detect single stars. We assume that the distribution of SHS is isotropic, an assumption that is increasingly valid for increasing $v_{\infty}$; as an example, the distance traveled by an SHS that was ejected $10^{10}$ yr ago exceeds 50 Mpc for $v_{\infty} \simeq 5 \times 10^{3}$~km~s$^{-1}$. SHS emitted with velocities lower than this value are likely to be somewhat concentrated toward the massive galaxies within which the majority are produced, we do not consider this effect here.

Initial velocities are drawn from the probability distribution displayed in Figure~\ref{fig:histogram}, but as previously stated, these velocities are only relative to the binary black hole system, and do not account for deceleration in galactic potentials. To determine the actual velocity at infinity, individual stars are assigned to individual merger events with a probability proportional to the secondary stellar mass in each merger. Then, the combined nuclear stellar masses and combined halo masses are used to determine the additional energy each SHS requires to escape its host galaxy. To determine the additional energy required to escape the nuclear cluster, we assume that the black hole dominates interior to the sphere of influence, and that the additional energy required to escape the nuclear cluster can be determined from the cluster's velocity dispersion, $v_{\rm nuc,esc} = 2 \sigma$, where $\sigma$ is defined as in Equation (\ref{eq:sigma}). We then use a relation for the maximum circular velocity of a halo \citep{Klypin:2011a}, multiplied by $\sqrt{2}$, to define the escape velocity from the merged halo,
\begin{equation}
v_{\rm halo,esc} = 930 \left(\frac{{\cal M}_{1} + {\cal M}_{2}}{10^{14} M_{\odot}}\right)^{0.316} {\rm km~s}^{-1},
\end{equation}
where we have adopted a Hubble parameter $h = 0.7$.

\begin{figure}
\centering\includegraphics[width=\linewidth,clip=true]{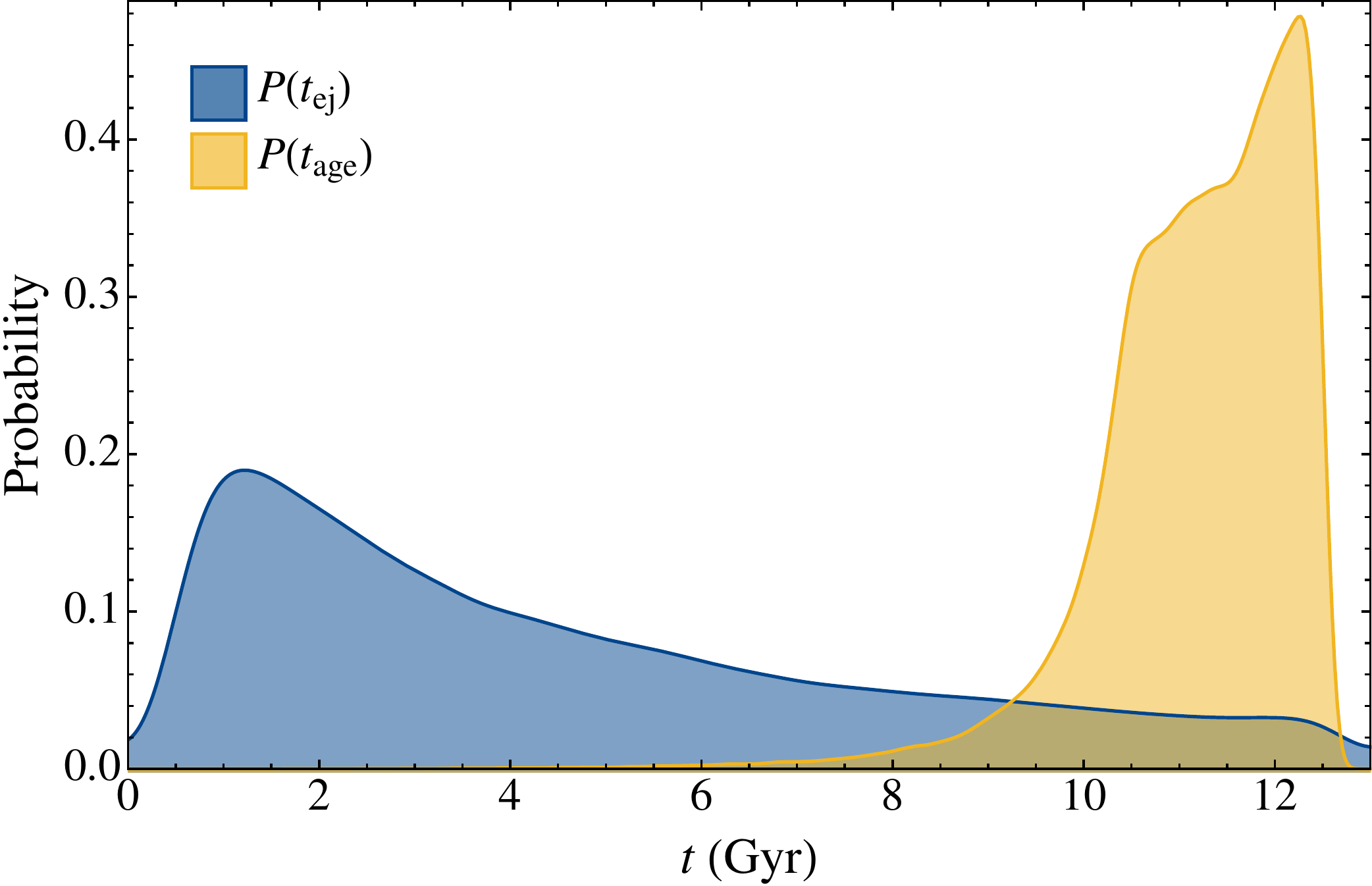}
\caption{Probability distribution of times of ejection $t_{\rm ej}$ (blue) and ages $t_{\rm age}$ (gold) of SHS. The derivation of these distributions is described in the text.}
\label{fig:ages}
\end{figure}

Now that the SHS's velocity upon leaving its origin galaxy is determined, there are two cosmological effects that must be taken into account to calculate their apparent velocity at the time of observation. For any freely-moving particle that is not bound to any structure, the momentum of the particle is reduced by a factor $a_{\rm ej}$, where $a_{\rm ej} = (1 + z_{\rm ej})^{-1}$ is the cosmological scale factor at the time the star was ejected \citep{Peebles:1993a,Weinberg:2008a}, where $z_{\rm ej}$ is the redshift of the source galaxy. This effect is quite significant for SHS as a significant fraction of merger activity occurs at high redshift (see Equation (\ref{eq:mergerrate})). This makes SHS rather unique: aside from individual particles such as photons, neutrinos, and cosmic rays, SHS are the only objects that are not bound gravitationally, and thus are potentially useful as sub-relativistic probes of cosmology, the details of how these stars may be used in this context are presented in a companion paper \citep{Loeb:2014a}. Secondly, stars a given distance from the MW will follow Hubble's law, giving them an additional component of recession velocity. This has the effect of making blueshifted SHS appear to move more slowly and redshifted SHS more quickly than an SHS at zero distance from the MW. The effect is relatively minor for the majority of SHS that are only a few Mpc from the MW, but in principle can be significant for bright stars that are visible out to the distance of Virgo, where the Hubble velocity is $\sim 10^{3}$~km~s$^{-1}$.

Taking these effects into account, the velocity of an SHS $\vec{v}_{\rm SHS}$ is then given by
\begin{align}
\vec{v}_{\rm SHS} &= v_{\rm H} \hat{r} + \vec{v}_{\rm p}\label{eq:vshs}\\
\vec{v}_{p} &= \frac{\vec{v}_{\infty}}{\left|\vec{v}_{\infty}\right|} \frac{\sqrt{\left|\vec{v}_{\infty}\right|^{2} - v_{\rm halo,esc}^{2} - v_{\rm nuc,esc}^{2}}}{1+z_{\rm ej}},
\end{align}
where $\vec{v}_{\rm p}$ is the star's peculiar velocity relative to the Hubble flow and where the Hubble velocity $v_{\rm H} = H_{0} r$ for $r \ll c/H_{0}$. Stars for which $\left|\vec{v}_{\rm p}\right|$ evaluates to an imaginary number are considered to remain bound to their host halos and are removed from the sample.

Given the velocity distribution of nearby SHS, the age of each star must be derived to determine what stage of its life it is in when it reaches us. SHS are composed of stars that were in existence at the time of each black hole merger, and therefore the redshift of formation $z_{\rm form}$ of a given SHS should be randomly drawn using the star formation history of the galaxy up to the redshift of ejection $z_{\rm ej}$. The time of ejection is determined by randomly drawing redshift values in proportion to the growth rate as a function of redshift, given in Equation (\ref{eq:mediangrowth}), where we consider a halo mass of $10^{12} M_{\odot}$ for simplicity (the difference in the relative growth rate is minuscule for different halo masses, scaling as ${\cal M}^{0.1}$). For the star formation history as a function of $z$ we use the dust-corrected star formation rates presented in \citet{Bouwens:2012a}, and draw $z_{\rm form}$ between $z_{\rm ej}$ and $z_{\max}$, where we set $z_{\max} = 10$. The resulting age distribution is shown in Figure~\ref{fig:ages}. Because most stars form at $z \sim 2$, the typical age of an SHS is $\gtrsim 10^{10}$~yr.

Once the local velocity distribution and age are determined, the distribution of stars that can be detected by a given survey can be determined by assigning a mass to each star and then using a stellar evolution track to evaluate its luminosity and color for its age, for this purpose we use the \mbox{PARSEC} 1.2S isochrones \citep{Bressan:2012a,Chen:2014a} available through the web-based tool CMD 2.7\footnote{\url{http://stev.oapd.inaf.it/cgi-bin/cmd}}. Again, we adopt a Kroupa mass function of stars. Given a mass, if a star's age is found to be greater than the maximum age within the PARSEC track corresponding to that mass (which do not include post-giant phases), we assume that the star has become a compact object, where stars with $M_{\ast} < 8 M_{\odot}$ become white dwarfs, $8 M_{\odot} < M_{\ast} < 20 M_{\odot}$ become neutron stars, and $M_{\ast} > 20 M_{\odot}$ become black holes.

\begin{figure*}
\centering\includegraphics[width=0.7\linewidth,clip=true]{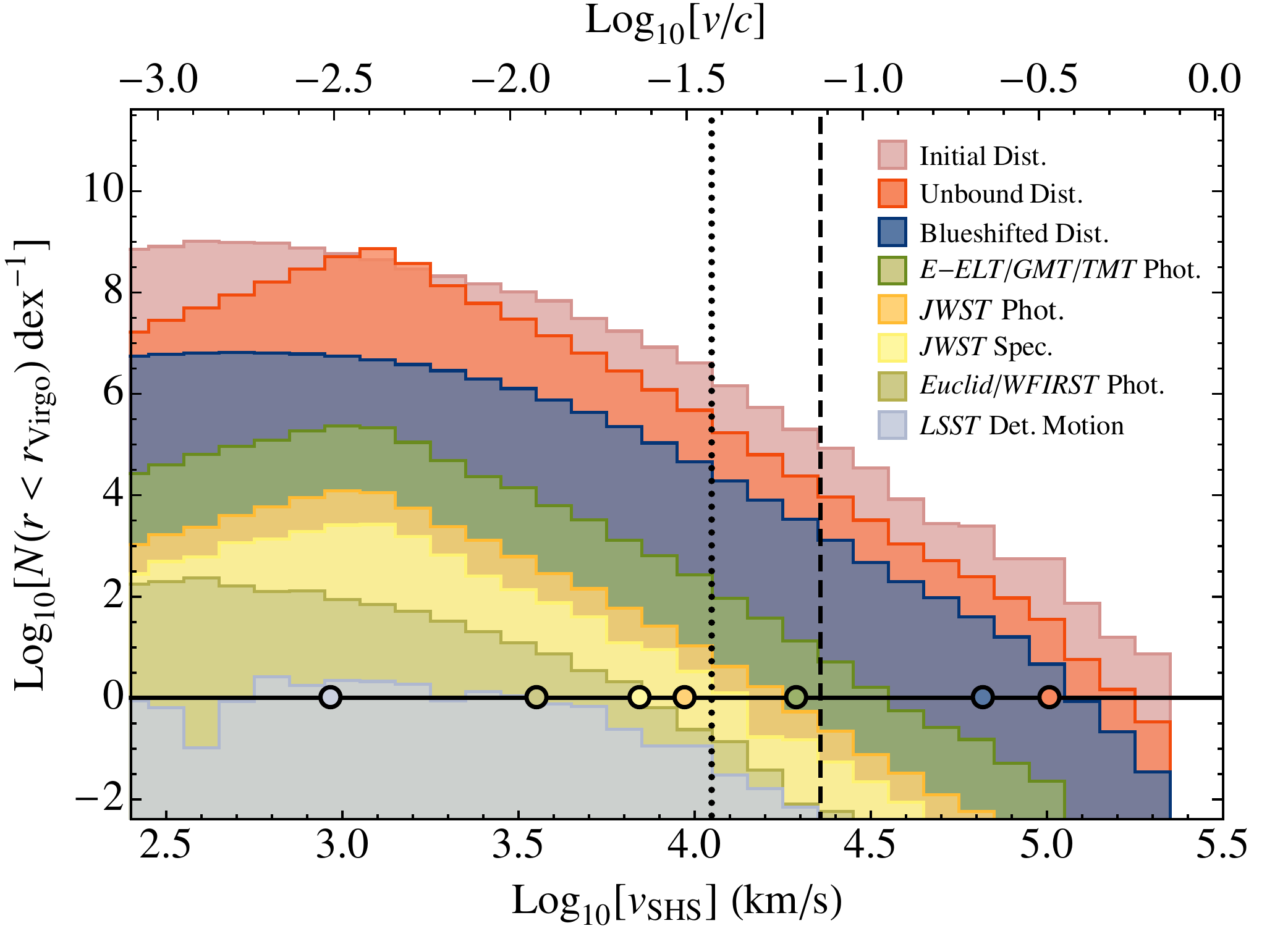}
\caption{Number of SHS satisfying various criteria at a distance $r$ less than that to the Virgo cluster ($r_{\rm Virgo} = 16$~Mpc). The pink histogram shows the initial distribution of SHS resulting from MBH mergers, where we have multiplied the total by a factor 2 to account for both primary and secondary SHS (see Section~\ref{sec:similarity}), and only considers the gravitational influence of the black holes themselves. The orange-red histogram shows the population of unbound SHS after accounting for the fraction of SHS that do not move fast enough to escape their host halos, resulting in a significant depletion of SHS for $v_{\infty} \lesssim 10^{3}$ km~s$^{\smash{-1}}$, and also accounts for the slowdown due to the expansion of the universe. The blue histogram only shows those stars that show a blueshift relative to the MW. The remaining histograms show the distributions expected to be detected by the various listed surveys, the assumed parameters of which are described in the text. The vertical dotted black line shows $v_{\infty} =$~11,000~km~s$^{\smash{-1}}$, the speed limit for the HVS mechanism with a binary star system and a single MBH, whereas the vertical dashed black line shows $v_{\infty} =$~24,000~km~s$^{\smash{-1}}$, the speed limit for HVS produced from SBH-star systems, all objects to the right of this line can only be produced by merging MBHs (see Section~\ref{sec:stellmasssecond}). The solid black line shows where the number count of objects within a specified distribution equals one, whereas the colored points along this line show the velocity for which the probability of detection for velocities greater than the marked value exceeds unity.}
\label{fig:histogram2}
\end{figure*}

The resulting distributions of stars that lie within a volume constrained to the distance of the Virgo cluster $r_{\rm Virgo} = 16$~Mpc is shown in Figure~\ref{fig:histogram2}. The pink histogram reproduces the histogram presented in Figure~\ref{fig:histogram}, where only the black hole's potential is taken into consideration, whereas the orange-red histogram accounts for the slowdowns associated with the nuclear cluster, host galaxy, cosmology, and redshift, as specified by Equation (\ref{eq:vshs}). Interestingly, this Figure~shows that the fastest star out to $r_{\rm Virgo}$ is likely moving with $v_{\rm SHS} \simeq c/3$, which translates to a Lorentz factor $\gamma = 1.06$. While the distribution does extend to these extreme velocities, the vast majority travel at a more pedestrian value of 1,000~km~s$^{-1}$, similar to the escape velocities of their host galaxies. Of these stars, less than half are blueshifted toward the milky way due to the redshift associated with the Hubble flow.

We consider several different kinds of future astronomical surveys and the yields of SHS that they are expected to detect, these estimates are largely based on the proposed capabilities for each survey and are only intended to roughly estimate the detected SHS distributions. For {\it LSST}, we consider SHS that satisfy both a magnitude and proper motion cut; stars must have I-band magnitudes < 25, with proper motion in excess of the limits presented in \citet{Ivezic:2008a}, $\sim 1$~mas~yr$^{-1}$. We include two all-sky spaced-based IR surveys of roughly similar capability, {\it Euclid} \citep{Laureijs:2011a} and {\it WFIRST} \citep{Green:2012a}, for both we require  an SHS to have H-band magnitude < 24 to be detected with no proper motion constraint, and assume that both surveys will cover $\sim 40\%$ of the sky. While {\it JWST}'s instruments will have very small fields of view (a few arcminutes), it will image some parts of the sky very deeply, potentially enabling the detection of SHS at the distance of Virgo; for this survey we presume that {\it JWST} will be in operation for 10 years and that any SHS bright enough to collected photometrically (K < 29) or spectroscopically (K < 26) would be considered detected. The next generation of large telescopes such as {\it E-ELT}\footnote{\url{http://www.eso.org/public/usa/teles-instr/e-elt}} \citep{Gilmozzi:2007a}, {\it GMT}\footnote{\url{http://www.gmto.org}} \citep{Johns:2012a}, and {\it TMT}\footnote{\url{http://www.tmt.org}} \citep{Macintosh:2006a} are also likely to have slightly larger fields of view (but still sub-degree) for their instruments; we similarly assume that these instruments will be in operation for a decade and that objects with K < 29 are detected.

We find that the fastest star in the volume out to Virgo likely travels with $v_{\infty} \simeq$~100,000~km~s$^{-1}$, with the fastest blueshifted star traveling with velocity 70,000 km~s$^{-1}$. With the above survey parameters in mind, the fastest star detectable with {\it E-ELT}/{\it GMT}/{\it TMT} travels with $v_{\infty} \simeq$~20,000~km~s$^{-1}$, the fastest star detectable with {\it JWST} travels with $v_{\infty} \simeq$~9,000~km~s$^{-1}$, the fastest star for which {\it JWST} spectroscopy is possible travels with $v_{\infty} \simeq$~7,000~km~s$^{-1}$, the fastest star likely to be detected with {\it Euclid/WFIRST} travels at 4,000~km~s$^{-1}$, and the fastest star that will have detectable proper motion in {\it LSST} travels with $v_{\infty} \simeq$~1,000~km~s$^{-1}$.

Figure~\ref{fig:stage} shows the stages for stars that are detectable with various surveys. Because the majority of stars in the universe have masses less than a solar mass, a large fraction of stars have yet to evolve off the MS. Around 10\% of stars have evolved beyond the giant branches to become a white dwarf, neutron star, or black hole, with most of these objects being white dwarfs (left-most column of Figure~\ref{fig:stage}). The remaining $\sim 1$\% are luminous giants, these are the stars that dominate the samples of surveys given the advanced age of SHS that lie near the MW (four right columns of Figure~\ref{fig:stage}). Red giant branch stars contribute about half of the detectable sample, with the shorter-lived (but brighter) giant branches contributing a significant fraction of the total. \citet{Palladino:2012a} identified a sample of candidate red giants with distances of 300 kpc -- 2 Mpc within SDSS and suggested that some of these stars may have a hypervelocity origin. While a large fraction of these stars are likely halo stars \citep{Bochanski:2014a}, these candidates are at distances for which we expect the SHS population to dominate HVS produced by the MW (see Figure \ref{fig:histogram}).

The fact that these evolved stages that are detectable only comprise a small fraction of the total population means that the number of detectable SHS is significantly reduced relative to the total number of SHS in the vicinity of the MW. Figure~\ref{fig:histogram2} also shows the expected velocity distributions of SHS within each survey's sample of detected SHS, the maximum velocity that is likely to be detected is $\sim$~20,000~km~s$^{-1}$, about a factor of five slower than the fastest star expected within $r_{\rm Virgo}$.

\section{Discussion}\label{sec:discussion}

\subsection{Detection versus Identification}\label{sec:ident}
While a number of SHS are likely to be detected by future surveys, definitive identification of SHS from photometry alone is likely to be challenging. Indeed, there will be many red, point-like objects in any selected imaging field, not just from MW dwarfs, but also from unresolved high-redshift background galaxies. This high chance of confusion is likely to hamper efforts to identify SHS that are stationary on the sky, i.e. those that are far from the MW.

The most straight-forward means of identification is to look for objects with high proper motions, and this will be probably doable with {\it LSST}. As described in the previous section, the vast majority of detectable SHS, regardless of the survey, are likely to be detected having velocities on the order of a few thousand km~s$^{-1}$. This means that their velocities are in fact quite similar to that of HVS produced by our own galaxy, or by nearby galaxies such as Andromeda \citep{Sherwin:2008a}. For these objects in which the velocities are similar to local HVS, the main identifying feature will be the fact that their velocity vectors are unlikely to originate from either our galactic center or Andromeda. Additionally, SHS are likely to outnumber the local HVS population at distances greater than a Mpc from the MW (see Figure~\ref{fig:histogram}).

\begin{figure}
\centering\includegraphics[width=\linewidth,clip=true]{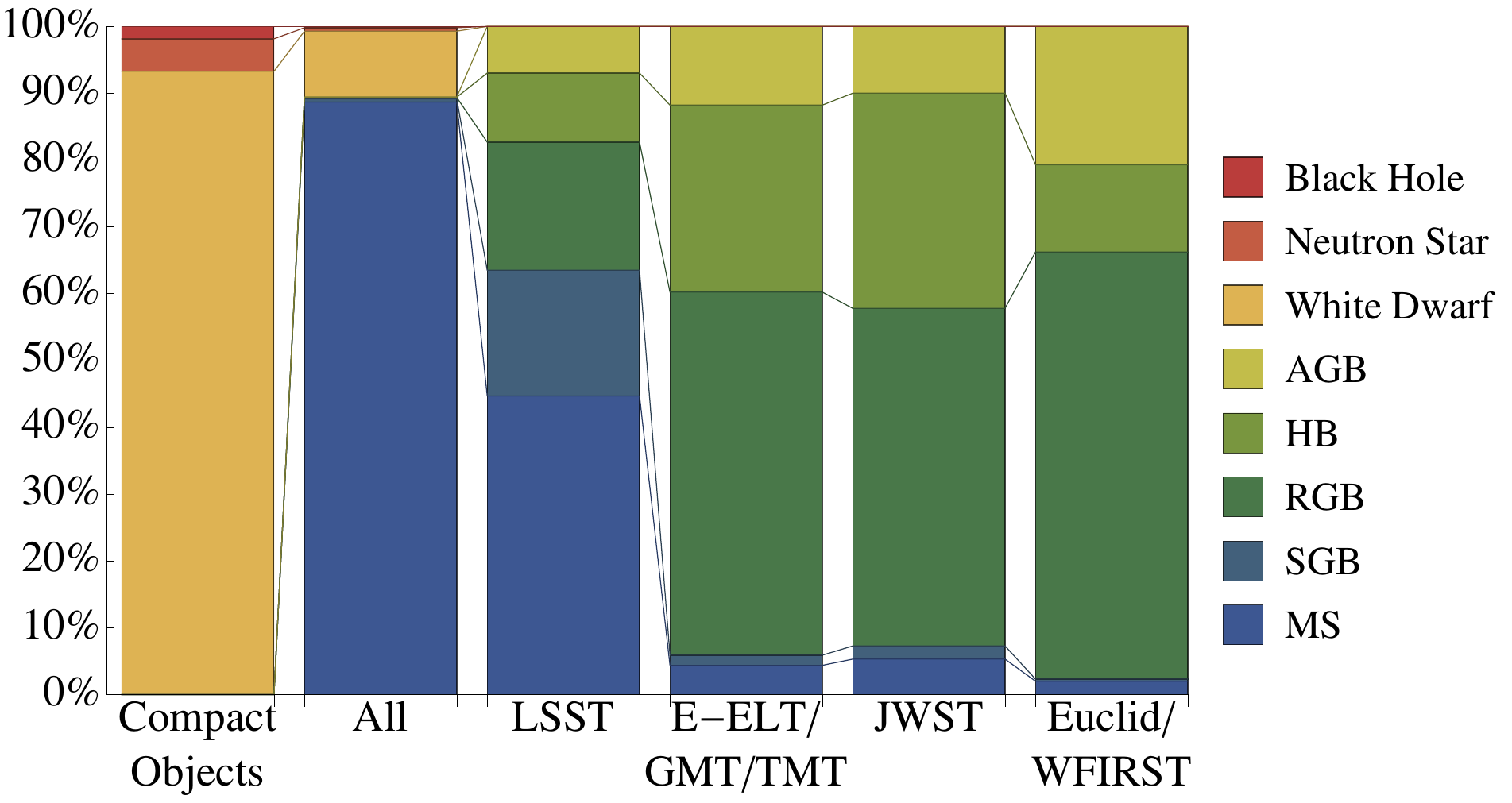}
\caption{Fraction of stellar types existent within a sphere with radius $r_{\rm Virgo}$ detectable with various telescopes. The first and second columns show the population of objects within this sphere, regardless of detectability, with the first column only showing the compact objects, and the second column showing all stellar types, which is dominated by low-mass stars that are still undergoing gravitational contraction. The subsequent four columns show the fraction of stellar types that are likely to be detected photometrically by various telescopes. The vast majority of detectable objects are evolved stars.\smallskip}
\label{fig:stage}
\end{figure}

However, we showed in Section~\ref{sec:beaming} that if the merger is plunging, many of the stars originally bound to the BMBH are likely to be beamed into a narrow cone. If this is the case for most BMBH, it would imply that the local density of SHS likely originates from only a couple BMBH mergers, and that several SHS would have trajectories that would point back to a common source galaxy \citep{Holley-Bockelmann:2005a}. In Figure~\ref{fig:m87} we assumed that the central Virgo cluster galaxy M87 experienced a BMBH merger 3 Gyr prior, where a large fraction of that event's SHS were beamed toward the MW. In such a scenario the trajectories of the SHS point back toward M87, potentially enabling its identification as the source galaxy. Because M87 is relatively close to us, all stars launched in this hypothetical merger event that are moving faster than 5,900~km~s$^{-1}$ have already passed beyond the 2 Mpc distance limit set in that figure, whereas those stars that are moving slower than 4,600~km~s$^{-1}$ have yet to reach us. This limited range in velocities, in conjunction with the spatial distribution of SHS and their velocity vectors, can potentially be used to associate families of SHS with particular galaxies and merger events.

For stars that move faster than a few thousand km~s$^{-1}$, their rarity means that their distance from the MW can be large enough that proper motion will not be detectable. For these stars, the best hope for identification is to happen to take a spectrum of an SHS. Because SHS have a high probability of being blueshifted toward the Milky Way (whereas most extended objects will be redshifted), a spectrum showing a blueshift on the order of a few thousand km~s$^{-1}$ or greater would permit their immediate identification. Figure~\ref{fig:specs} shows that the spectral shift associated with the fastest star that is likely to be detected by a thirty-meter class telescope would be extremely obvious. If only photometry is available, these stars will also have a measurable shift in color on the order of a few tenths of a magnitude, however given that these stars are most likely to be giant branch stars, this small shift in color is likely difficult to disentangle from color changes associated with normal giant evolution.

\begin{figure}
\centering\includegraphics[width=0.9\linewidth,clip=true]{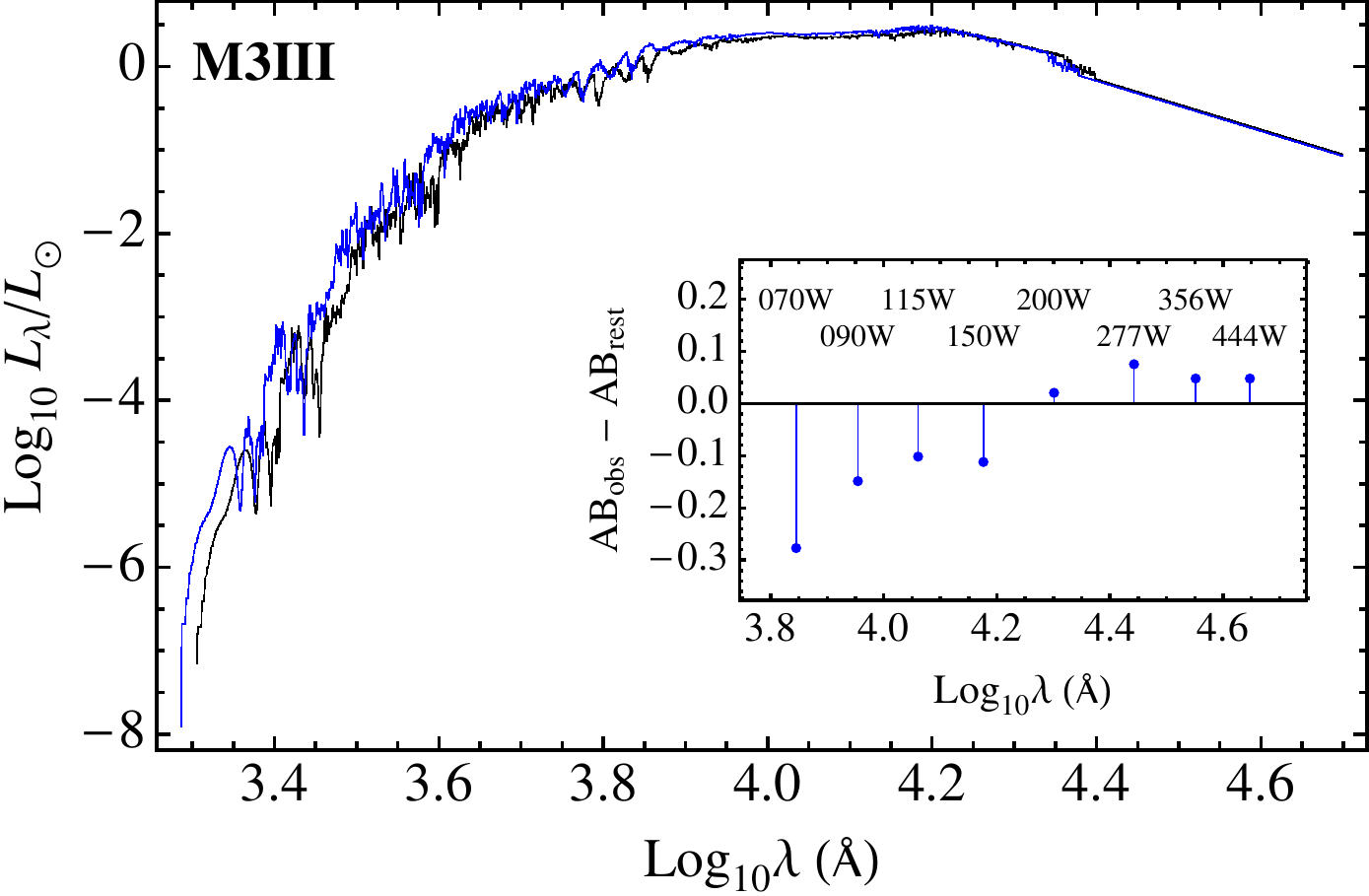}
\caption{Comparison of rest-frame red giant spectra (in black) to blueshifted spectra (in blue) corresponding to the fastest star likely to be detected by {\it E-ELT}/{\it GMT}/{\it TMT}, $v_{\rm SHS}~=~0.06c \simeq$~20,000 km s$^{\smash{-1}}$. The rest-frame spectrum is drawn from the publicly available library of \citet{Pickles:1998a} and shows a M3III giant. As the input spectra only extend to \smash{25,000~\AA}, fluxes redward of this value are extrapolated assuming a Rayleigh-Jeans law. Inset within the figure is the difference between the observed AB magnitude AB$_{\rm obs}$ for individual {\it JWST} filters versus the magnitude AB$_{\rm rest}$ that would be observed if the star were at rest.}
\label{fig:specs}
\end{figure}

\subsection{Other Means of Detection}
So far we have only considered single, luminous stars in our analysis as these are likely to be the most common SHS. However, there are several other ways to either detect them directly or to infer their presence. We describe a few possibilities in this section.

\subsubsection{Accreting and Merging SHB}
As described in Section~\ref{sec:shb}, the mechanism for accelerating SHS can also produce SHB. These binaries are likely to have short orbital periods (otherwise they would not have survived the acceleration process), and thus are likely to eventually undergo mass transfer, followed potentially by a merger. If one of the two stars is a compact object this mass transfer can illuminate the binary to the Eddington limit ($\sim 10^{38}$~ergs~s$^{-1}$ for solar mass accretor), potentially making the system visible to very large distances. Likewise, if the two stars merge around the time they are close to the MW, they may form a blue straggler with a luminosity comparable to a giant that would potentially be more detectable with optical surveys. However, both X-ray binaries and blue stragglers are rather rare in the field, and given that generic SHB consisting of two MS stars are rare in the first place, it implies that SHB would have to preferentially form such systems to have an appreciable probability of detection.

\subsubsection{Bow Shocks}
An additional means of detection may be via the bow shocks driven into the ISM by these high-velocity stars. It is already known that fast-moving stars in star-forming regions can drive very strong and spatially large bow-shocks into their environments \citep{Meyer:2014a}, and thus this may be a promising way to detect the presence of a fast-moving star. However, the SHS are unlikely to be significantly concentrated toward the galactic plane where the densest phases of the ISM exist \citep[although there may be a local density enhancement of $\sim 100$ for lower-velocity SHS due to gravitational focusing, see][]{Sherwin:2008a}, and therefore their bow shock signatures are most likely to be confined to regions of fairly low density, resulting in luminosities that are unlikely to exceed the luminosities of the stars themselves.

\subsubsection{Isolated Local Group Pulsars}
While neutron stars are likely to only be $\lesssim 1\%$ of all SHS, this means that roughly $10^{3}$ would lie within 1 Mpc of the MW. Some fraction of these objects may be pulsars that are potentially detectable with SKA \citep{Smits:2009a,Lazio:2013a}.

\begin{figure*}
\centering\includegraphics[width=0.8\linewidth,clip=true]{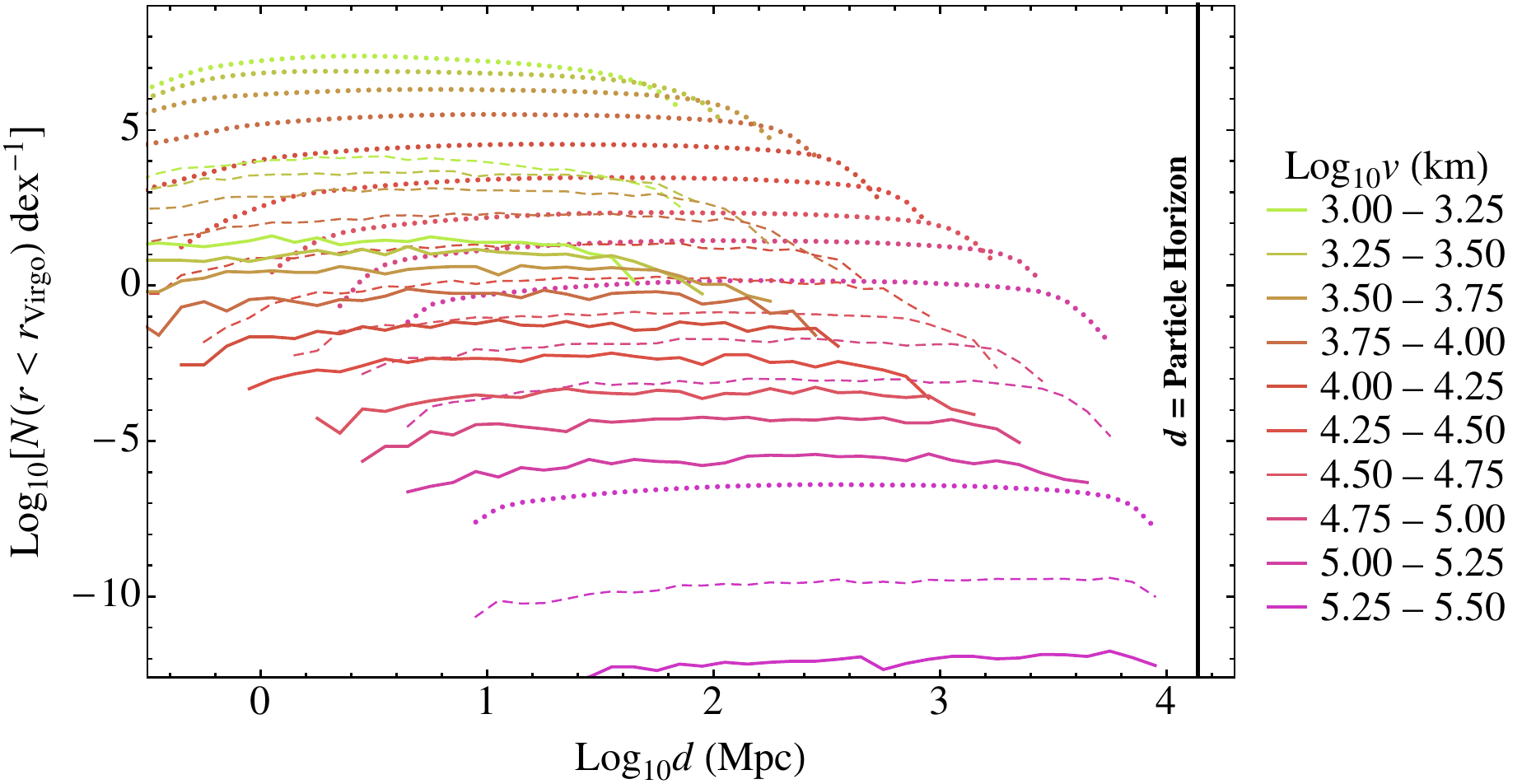}
\caption{Histograms of distance $d$ traveled by SHS from their point of origin for SHS moving at various velocities. The different colored lines show different velocity ranges (as labeled), while the line-styles refer to different criteria for detection: The dotted lines show all objects, the dashed lines show objects detectable photometrically with {\it JWST}, and the solid lines show objects detectable with {\it Euclid}/{\it WFIRST}. The sharp drop-off in the last bin shows where a large fraction of stars are swallowed whole rather than being ejected as SHS.\vspace{3em}}
\label{fig:traveld}
\end{figure*}

\subsubsection{Intrahalo Light (IHL)}
There are about $10^{10}$~Mpc$^{-3}$ stars in the local universe, and with the SHS density being approximately $10^{5}$~Mpc$^{-3}$, SHS make up about $10^{-5}$ of all stars in the universe at $z = 0$. Because they travel great distances from their source galaxies (Figure~\ref{fig:traveld}), SHS are expected to have a nearly isotropic distribution, and thus will also occupy the vast voids between galaxy clusters that are otherwise unlikely to host many stars. This population may be detectable via their contribution to IHL, which we estimate to yield $\sim 10^{-3}$~nw~m$^{-2}$~sr$^{-1}$ at 1 micron. This is lower than the average IHL value of $\sim 1$~nw~m$^{-2}$~sr$^{-1}$ \citep{Cooray:2012b}, but is comparable to the minimum value of $\sim 10^{-3}$~nw~m$^{-2}$~sr$^{-1}$ seen in some regions of the sky \citep{Zemcov:2014a}. This suggests that SHS may indeed set the floor value for the IHL in voids.

\subsubsection{Supernovae from SHS and SHB}
Far-flung supernovae that occur in no known host galaxy are potentially another way of inferring the presence of SHS and SHB. For SNe II, which occur only for stars with masses greater than $\sim 8 M_{\odot}$, the range of the ejection distances is limited, but potentially can be distinguished. An $8 M_{\odot}$ star traveling at 3,000~km~s$^{-1}$ will travel at most 300 kpc before death, at these distances it is possible that the supernovae occurred in a unresolved dwarf galaxy. Given the local density of SHS and presuming that the per-star supernova rate is the same as the MW ($\sim 10^{-4}$~Mpc$^{-3}$~yr$^{-1}$), $\sim 100$~yr$^{-1}$ SNe II will occur in the volume out to $z \sim 1$ at these characteristic distances (compare this to the total Type II rate within that volume of $\sim 10^{7}$~yr$^{-1}$). Normal HVS can also contribute to these off-center Type II events as they eject stars at similar speeds. For stars moving at 30,000~km~s$^{-1}$, about one Type II will occur at a distance $\sim 3$ Mpc from its host per century out to $z = 1$.

For SNe Ia, the rates are reduced somewhat by the fact that the progenitors are binary stars, of which an unknown fraction survives the ejection process (see Section~\ref{sec:shb}). By definition, high velocity Type Ia progenitors {\it cannot} be generated by the HVS process, which only accelerates single stars, and instead must either come from a triple system disruption or the SHB mechanism described in this paper. A key difference between these systems and Type II progenitors is that they can potentially live much longer before the supernova occurs, and thus can be much further from their source galaxies. If double-degenerate systems are responsible for a significant fraction of Ia events, or the ejection occurs before the donor star evolves off the MS in the single-degenerate phase (to avoid having a fragile giant in the SHB), their travel distance is only limited by the Hubble time, and thus these supernovae can occur in truly isolated environments, perhaps even at the centers of voids where very few stars reside.

Given that LSST is expected to detect a few $10^{5}$ supernovae per year \citep{Abell:2009a}, and presuming the above rates, we predict that LSST should find a few isolated supernovae (with $d \gtrsim 100$~kpc from their origin galaxy) that originate from the SHS/SHB population per year. \citet{Zinn:2011a} found several candidate supernovae at distances of kpc from their source galaxies that likely originated from stars with very short progenitor lifetimes. They inferred that the progenitor stars must have traveled with speeds in excess of several hundred km~s$^{\smash{-1}}$ to get so far away from their source galaxies. However, it should be noted that a contaminating foreground of both Type Ia and Type II would exist in galaxy clusters due to tidal stripping of stars from member galaxies \citep{Maoz:2005a}. Therefore, the best place to look for supernovae originating from SHS/SHB are around galaxies that are not in clusters, or in the voids between galaxies.

\subsection{Complications}
Aside from checking whether stars cross the IBCO of either black hole, our scattering experiments do not include the effects of general relativity, which can be quite large for those stars that we predict are launched with velocities of $c/3$. While including these effects is likely to affect the outcome of an individual scattering experiment, we think its inclusion is unlikely to greatly affect the resulting statistics of ejection velocities. As shown in Figure \ref{fig:fate}, the fraction of stars that come with a few IBCO radii of either black hole and avoid being swallowed or tidally disrupted is quite small.

We assumed that all merging MBHs within the mass ratio range specified merge eccentrically, and that the evolution of the secondary's orbit is plunging. While both numerical simulation and Fokker-Planck approaches have found that large eccentricities can be excited, net rotation of the primary's nuclear cluster can suppress the build-up of eccentricity for as many as half of all systems \citep{Sesana:2011a,Dotti:2012a}.

For the fastest SHS ($v > 10^{4}$~km~s$^{-1}$), the velocity distribution is likely quite sensitive to the radial distribution of stars interior to the distance at which star-star collisions become common (see Section \ref{sec:experiments}). In this paper we assumed that this distribution is defined based on rather limited observations of the MW's nuclear cluster, but this observed distribution is not necessarily generally applicable to the $\sim 10^{8} M_{\odot}$ black holes responsible for producing the majority of SHS. Because two-body relaxation is ineffective at replenishing stars interior to the radius where collisions become common \citep[The results of which would produce luminous transients, see][]{Balberg:2013a}, it is largely resonant interactions \citep{Hamers:2014a} and binary disruptions \citep{Perets:2009b} that populate this region. Additionally, the secondary's nuclear cluster is continuously subjected to very strong perturbations from the primary that are likely to alter the orbits of all stars within it, even if they do not immediately unbind.

\section{Conclusions}\label{sec:conclusion}
We have demonstrated in this paper that SHS (and SHB) are stars that can be accelerated to speeds in excess of $10^{4}$ km~s$^{-1}$, a speed which is difficult (if not impossible) to produce via any other astrophysical mechanism. If discovered, they would be unique tracers of the eccentric mergers of supermassive black holes. There is some observational evidence that such mergers do occur \citep{Valtonen:2008a,Batcheldor:2010a}, but this evidence will likely remain inconclusive until gravitational waves are detected from them. Because the large velocities found here are contingent upon merging supermassive black holes possessing significant eccentricities, the discovery of even one SHS or SHB would suggest that eccentric MBH mergers are common.

While identifying these stars may be challenging, we have shown that many of them are likely to be detected by future astrophysical surveys, and that they may be discovered via several other direct and indirect means.

These fastest stars are also one of the few natural phenomena that are likely to cross the vast chasms of empty space between galaxies.  Figure~\ref{fig:traveld} shows that even the slowest SHS, which move at a few $10^{3}$~km~s$^{-1}$, will have traveled tens of Mpc from their source galaxy by the time they reach the MW, and that the fastest, although quite rare, can travel nearly 10 Gpc since being ejected. This makes them potentially powerful probes of cosmological expansion, which we detail in a companion paper \citep{Loeb:2014a}. Given the rates of production we have calculated here, it is very likely that a star that has traveled a distance of over 1 Gpc lies between the MW and the Virgo cluster. Such a star would be by far the fastest, and most-traveled, luminous object in our local neighborhood.

\acknowledgements
We are thankful for fruitful discussions with F.~Antonini, K.~Batygin, A.~Bogdan, W.~Brown, C.~Conroy, S.~Genel, I.~Ginsburg, P.~Groot, M.~Holman, S.~Naoz, E.~Ramirez-Ruiz, R.~Sari, A.~Sesana, D.~Sijacki, J.~Strader, and Y.~Levin. We are especially grateful to M.~C.~Miller for extended discussions regarding the mechanism presented here, and to our referee who provided a more direct derivation of the maximum speed for hypervelocity stars. This work was supported by Einstein grant PF3-140108 (J.~G.) and NSF grant AST-1312034 (A.~L.). Much of this paper was written at the Aspen Center for Physics (NSF Grant \#1066293), who we thank for their generous hospitality.

\bibliographystyle{apj}
\bibliography{/Users/james/Dropbox/library}

\end{document}